\documentclass[twocolumn,showpacs,preprintnumbers, 
               superscriptaddress,nofootinbib]{revtex4-1}
\usepackage{graphicx, fancybox}
\usepackage{amsmath,amssymb,bm}

\begin{document}

\title{
  Dynamical evolution of critical fluctuations and
  its observation in heavy ion collisions
}

\author{Miki Sakaida}
\email{sakaida@kern.phys.sci.osaka-u.ac.jp}
\affiliation{
Department of Physics, Osaka University, Toyonaka, Osaka 560-0043, Japan}

\author{Masayuki Asakawa}
\email{yuki@phys.sci.osaka-u.ac.jp}
\affiliation{
Department of Physics, Osaka University, Toyonaka, Osaka 560-0043, Japan}

\author{Hirotsugu Fujii}
\email{hfujii@phys.c.u-tokyo.ac.jp}
\affiliation{
	Institute of Physics, University of Tokyo, Tokyo 153-8092, Japan}

\author{Masakiyo Kitazawa}
\email{kitazawa@phys.sci.osaka-u.ac.jp}
\affiliation{
Department of Physics, Osaka University, Toyonaka, Osaka 560-0043, Japan}
\affiliation{
J-PARC Branch, KEK Theory Center,
Institute of Particle and Nuclear Studies, KEK,
203-1, Shirakata, Tokai, Ibaraki, 319-1106, Japan }

\begin{abstract}

We study time evolution of critical fluctuations of conserved charges 
near the QCD critical point 
in the context of relativistic heavy ion collisions.
A stochastic diffusion equation is employed in order to describe 
the diffusion property of the critical fluctuation arising from 
the coupling of the order parameter field to conserved charges.
We show that the diffusion property 
gives rise to a possibility of probing the early time fluctuations 
through the rapidity  window dependence
of the second-order cumulant and correlation function of conserved charges.
It is pointed out that their non-monotonic behaviors 
as functions of the rapidity interval are robust experimental signals 
for the existence of the critical enhancement around
the QCD critical point. 
\end{abstract}

\date{\today}
\preprint{J-PARC-TH-0088}

\maketitle

\section{Introduction}
\label{sec:intro}

The search for the QCD critical point is 
one of the most intriguing topics in the physics of QCD in medium.
The existence of the critical point(s)
in the QCD phase diagram in the 
temperature ($T$) and baryon chemical potential ($\mu$) plane 
is suggested by effective models \cite{Asakawa:1989bq,
Barducci:1989wi,Halasz:1998qr, Berges:1998rc, Scavenius:2000qd, Hatta:2002sj} 
and lattice QCD Monte Carlo simulations 
\cite{Fodor:2001pe,deForcrand:2002hgr,Endrodi:2011gv}. 
However, the number~\cite{Kitazawa:2002bc,Hatsuda:2006ps}, locations, 
and even existence itself are still controversial.
The experimental search for the critical point is one of the 
fundamental purposes of the relativistic heavy ion collision 
experiments~\cite{Luo:2017faz}.
Because the medium created by these experiments pursues 
different trajectories in the $T$--$\mu$ plane 
depending on collision energies $\sqrt{s_{_{NN}}}$, 
this search would be achieved by comparing the characteristics of 
collision events at various $\sqrt{s_{_{NN}}}$.
For this purpose, active experimental studies in the 
Beam-Energy Scan (BES) program 
at the Relativistic Heavy-Ion Collider (RHIC) are ongoing~\cite{BESII}.
Future heavy-ion program at J-PARC~\cite{J-PARC-HI}, FAIR~\cite{FAIR} 
and NICA~\cite{NICA} will also contribute to this subject.

In the experimental search for the QCD critical point,
fluctuation 
observables, such as the cumulants of the net-baryon number, 
are believed to be natural and
promising observables~\cite{Asakawa:2015ybt}.
At the critical point in the thermodynamic limit,
equilibrated fluctuations of various observables diverge 
reflecting the softening of the effective potential. 
Such singular behavior is 
expected to be found in event-by-event
analyses in heavy ion collisions
\cite{Stephanov:1999zu,Jeon:2003gk,Stephanov:2008qz,Asakawa:2009aj}.
The experimental analyses of the fluctuation observables, especially
focusing on non-Gaussianity, have been performed actively 
recently~\cite{Adamczyk:2013dal,Adamczyk:2014fia,Luo:2017faz}.

In interpreting the experimentally-observed fluctuations, however,
it should be remembered that the fluctuations observed in these 
experiments are not those in an equilibrated medium near the critical point.
First, the more closely the system approaches the critical point, the longer becomes
the relaxation time toward equilibrium because of the critical slowing down~\cite{Berdnikov:1999ph}.
Owing to this effect, the enhancement of fluctuations is limited 
even if the medium passes right through the QCD critical point.
Second, 
the critical fluctuations are to be distorted during the subsequent evolution
until the detection~\cite{Kitazawa:2013bta,Sakaida:2014pya}.
Remember that particles are measured only in the final state.
To extract the information on the critical point from the 
experimental data, proper understanding and description of 
these effects are indispensable, in addition to various
restrictions from the real experimental settings~\cite{Asakawa:2015ybt}.

The time evolution of critical fluctuations has been discussed in 
the literature~\cite{Berdnikov:1999ph,Nonaka:2004pg,Kapusta:2012zb,
Mukherjee:2015swa,Herold:2016uvv}.
In Refs.~\cite{Berdnikov:1999ph,Nonaka:2004pg,Mukherjee:2015swa},
the effect of the critical slowing down is discussed for 
fluctuation of a uniform order parameter field 
$\sigma=\langle\bar qq\rangle$. 
It is known, however, that the critical mode of the QCD critical point is 
not the pure $\sigma$ mode, but rather is given by
a linear mixing of the $\sigma$ mode  and conserved charges
\cite{Fujii:2003bz,Fujii:2004jt,Son:2004iv,Minami:2011un}.
The soft mode of the critical point thus is a diffusion mode.
As we will see below, the time evolution of a diffusion mode
depends on the length scale, and this property is 
crucial for the description of its dynamics.
One of the main goals of the present study is
to reveal the dynamical 
evolution of the critical fluctuation with the diffusion property.

Another aim of this paper is
to make a connection between the critical fluctuation
of thermodynamics and experimental observation.
This connection will be clearest for the fluctuations
of conserved charges, especially that of the net-baryon number
(see Ref.~\cite{Asakawa:2015ybt} for detailed discussions).
In order to understand the evolution of the conserved-charge fluctuations,
we first need to know that of the charge density itself
in a collision event.
This problem has been discussed for a hadronic medium in a simple setup 
\cite{Kitazawa:2013bta,Sakaida:2014pya},
but to the best of our knowledge it has not yet been studied for the case where
a critical enhancement is encountered during the evolution.
Carrying out this subject is the second goal of this study.

In this study, in order to address these issues
we employ a simple stochastic diffusion equation (SDE).
In this approach, the singularity associated with the critical point
is encoded in the time-dependent susceptibility and diffusion 
coefficient.
The SDE is a counterpart of the stochastic hydrodynamics
\cite{Kapusta:2011gt}, and is a good model to describe fluctuations of 
conserved charges.
Moreover, as shown in Refs.~\cite{Fujii:2003bz,Fujii:2004za,Fujii:2004jt,Son:2004iv},
the critical mode at the QCD critical point has the diffusion character
and should be described by the SDE. 

In this study, we first write down a formal solution
for the density fluctuation within rapidity window $\Delta y$.
Next, with a phenomenological parametrization for the 
susceptibility and diffusion coefficient,
we analyze numerically the time evolution of the fluctuation.
We find that the time evolution of the fluctuations strongly depends 
on the size of the rapidity window $\Delta y$.
We also show that the second-order cumulant can have a non-monotonic 
 $\Delta y$ dependence only when 
the medium fluctuation 
undergoes a critical enhancement during the time evolution.
It is argued that this non-monotonic behavior serves as 
a robust
experimental signal of the critical enhancement.

In the present study we restrict
our attention only to the second-order
cumulant, and leave the analysis of third and still
higher order cumulants for future study,
because even at the second order non-trivial outcomes are obtained
from the analysis.
We also deal with the correlation functions
of conserved charges.
We show that the same argument on the non-monotonic behavior
holds for this function with rapidity 
window replaced by rapidity separation.
The relation between the cumulant and the correlation function is also 
studied in detail.

This paper is organized as follows.
In the next section we introduce the SDE,
and argue that this model is suitable for describing the dynamics of
conserved-charge fluctuations including the critical ones.
We then solve this equation analytically, 
and discuss general properties in Sec.~\ref{sec:analytic}.
In Sec.~\ref{sec:time}, we introduce a phenomenological model, which 
is numerically solved in Sec.~\ref{sec:real}.
The last section is devoted to discussions and a summary.
In Appendix~\ref{sec:SDEc}, we give a brief review of 
Refs.~\cite{Fujii:2003bz,Fujii:2004jt,Son:2004iv}, which clarified 
the diffusion property of the critical fluctuation.
In Appendix~\ref{sec:discussion}, we compare the cumulant
and correlation function and discuss the condition for 
the appearance of the non-monotonic behaviors in these functions
in detail.

\section{Model}
\label{sec:model}

\subsection{Second-order cumulant and correlation function}
\label{subsec:<Q>}

In this study, we investigate the time evolution of the second-order cumulant 
and the correlation function of conserved charges in the hot medium 
produced in heavy ion collisions.
We assume a boost-invariant Bjorken model for the event evolution
throughout this paper in order to simplify the analysis,
and adopt the Milne coordinates, 
the spacetime rapidity 
$y=\tanh^{-1}(z/t)$ and proper time $\tau=\sqrt{t^2-z^2}$.

Let us consider a conserved-charge density per unit rapidity $n(y,\tau)$,
where transverse coordinates have been integrated out.
As the conserved charge,
we can take specifically the net-baryon number~\cite{Kitazawa:2011wh}
or net-electric charge in heavy ion collisions.
The amount of the charge in a finite rapidity interval $\Delta y$ at 
mid-rapidity at proper time $\tau$ is given by 
\begin{align}
Q_{\Delta y}(\tau) = \int_{-\Delta y/2}^{\Delta y/2} dy\, n(y,\tau).
\end{align}
The second-order cumulant, or variance,  of $Q_{\Delta y}(\tau)$ 
is written as
\begin{align}
\langle Q_{\Delta y}(\tau)^2 \rangle_{\rm c}
&= \langle \delta Q_{\Delta y}(\tau)^2 \rangle
\nonumber\\
&=\int_{-\Delta y/2}^{\Delta y/2} dy_1 dy_2 
\langle \delta n(y_1,\tau) \delta n(y_2,\tau) \rangle,
\label{eq:<Q^2>}
\end{align}
where $\langle \delta n(y_1,\tau) \delta n(y_2,\tau) \rangle$ 
is the correlation function with
$\delta n(y,\tau) = n(y,\tau) - \langle n(y,\tau) \rangle$ and
$\langle \cdot \rangle$ stands for event average.
In a boost invariant system, the correlation function depends 
only on the rapidity difference $\bar{y}=y_1-y_2$, and 
Eq.~(\ref{eq:<Q^2>}) is rewritten as 
\begin{align}
\langle Q_{\Delta y}(\tau)^2 \rangle_{\rm c}
= \int_{-\Delta y}^{\Delta y} d\bar{y} \, (\Delta y - |\bar{y}|) \,
\langle \delta n(\bar{y},\tau) \delta n(0,\tau) \rangle.
\label{eq:<Q^2>cor}
\end{align}
This reveals a close relation between 
$\langle Q_{\Delta y}(\tau)^2 \rangle_{\rm c}$ and 
$\langle \delta n(\bar{y},\tau) \delta n(0,\tau) \rangle$.
We investigate both of these functions in this paper.
From Eq.~(\ref{eq:<Q^2>cor}), one finds that 
\begin{align}
\lim_{\Delta y\to0} \frac{d~}{d \Delta y} 
\frac{\langle Q_{\Delta y}(\tau)^2 \rangle_{\rm c}}{\Delta y}
= \lim_{\bar{y}\to0} \langle \delta n(\bar{y},\tau) \delta n(0,\tau) \rangle.
\label{eq:Dy->0}
\end{align}

\subsection{Stochastic diffusion equation}
\label{subsec:SDE}

We consider the time evolution of
the conserved charge density $n(y,\tau)$
at long time and length scales.
This is well described by the stochastic diffusion equation
\cite{Kapusta:2011gt,Asakawa:2015ybt},
which is written in the $\tau$--$y$ coordinates as
\begin{eqnarray}
\label{eq:SDE}
\frac{\partial}{\partial \tau}\delta n(y,\tau)=D_y(\tau)
\frac{\partial^2}{\partial^2 y}\delta n(y,\tau)
+\frac{\partial}{\partial y} \xi(y,\tau)
\, ,
\end{eqnarray}
where the diffusion coefficient $D_y(\tau)$ 
is related to the Cartesian one 
$D_{\rm C}(\tau)$ as $D_y(\tau)=D_{\rm C}(\tau)\tau^{-2}$.
The noise term $\xi(y,\tau)$
represents the coupling with ``short-time'' fluctuations,
whose average should vanish $\langle\xi(y,\tau)\rangle=0$.
The noise term appears with the $y$-derivative in Eq.~(\ref{eq:SDE})
so as to satisfy the conservation constraint.

When the noise correlation is local in time and space,
the fluctuation-dissipation relation specifies its value to be 
\cite{Asakawa:2015ybt}
\begin{eqnarray}
\label{eq:noise}
\lefteqn{
  \langle\xi(y_1,\tau_1)\xi(y_2,\tau_2) \rangle }
\nonumber\\
&= 2 \chi_y(\tau)D_y(\tau) \delta(y_1-y_2)\delta(\tau_1-\tau_2).
\end{eqnarray}
Here, $\chi_y(\tau)$ denotes the susceptibility of the conserved charge 
per unit rapidity and is related to the one in the Cartesian coordinates 
$\chi_{\rm C}(\tau)$ as $\chi_y(\tau)/\tau=\chi_{\rm C}(\tau)$.
The susceptibility $\chi_y(\tau)$ is related to the second-order cumulant
in equilibrium as
$\langle Q_{\Delta y}^2 \rangle_{\rm c,eq}=\chi_y \Delta y$.
Furthermore, the noise is independent of the value of the
density at earlier times:
$\left <n(y,\tau) \xi(y',\tau')\right > =0$ for $\tau\le\tau'$.
In the rest of this paper, we suppress the subscripts of 
the diffusion coefficient and susceptibility, and denote 
$D(\tau)=D_y(\tau)$ and $\chi(\tau)=\chi_y(\tau)$, respectively.

The SDE~(\ref{eq:SDE}) is not only a suitable equation for describing the time 
evolution of conserved charges in a non-critical system,
but also a good phenomenological equation to deal with the slow dynamics
near the QCD critical point.
This is because there the critical mode is identified as a linear 
combination of the $\sigma$ mode and conserved charges, and therefore 
its evolution must be consistent with the conservation law
\cite{Fujii:2003bz,Fujii:2004jt,Son:2004iv,Kapusta:2012zb}.
As a consequence, the equation for the critical mode 
is given by the same form as in Eq.~(\ref{eq:SDE}).
See Appendix~\ref{sec:SDEc} for more detailed discussion.
Hence, in the present study we use Eq.~(\ref{eq:SDE}) solely to 
describe the time evolution throughout the event trajectories passing 
near and away from the critical point.

In our study, the critical enhancement and slowing down of fluctuations
are represented by the $\tau$-dependent susceptibility $\chi(\tau)$ 
and diffusion coefficient $D(\tau)$:
Near the critical point, $\chi(\tau)$ 
grows sharply and $D(\tau)$ becomes vanishingly small 
reflecting, respectively, the large fluctuation and critical 
slowing down.

In the next section, we derive the
formal solution of Eq.~(\ref{eq:SDE}) 
with $\tau$ dependent $\chi(\tau)$ and $D(\tau)$, and 
discuss its general property.
We then model the trajectories of collision events in terms of
possible $\tau$-dependence of the susceptibility $\chi(\tau)$ and
diffusion coefficient $D(\tau)$ including the effect of the critical
point, and analyze the time evolution numerically 
in Secs.~\ref{sec:time} and \ref{sec:real}.

\subsection{Comment on critical slowing down}

Before closing this section, 
we comment on the difference of the treatment of the critical
fluctuation in the present study from previous 
ones~\cite{Berdnikov:1999ph,Nonaka:2004pg,Mukherjee:2015swa}.
In Refs.~\cite{Berdnikov:1999ph,Nonaka:2004pg,Mukherjee:2015swa},
the time evolution of the uniform $\sigma$ field
was analyzed as the slowest mode of the system%
\footnote{
In Refs.~\cite{Berdnikov:1999ph,Nonaka:2004pg}, 
the evolution of the ``correlation length'' of the $\sigma$ mode 
is studied, which is equivalent to treating the
spatially uniform $\sigma$ mode,
as shown in Ref.~\cite{Mukherjee:2015swa}.}.
Since the $\sigma$ mode is non-conserved, it
can relax locally in rapidity space and 
follows a relaxation equation without $\partial/\partial y$.

This assumption is in contrast to the conserved charge
fluctuation discussed in the present study,
which can relax only through diffusion.
As discussed already, the evolution equation of the critical 
fluctuation has to be consistent with the conservation law
\cite{Fujii:2004za,Fujii:2004jt}.
In order to respect this property, the evolution has to be dealt 
with the equations like the SDE consistent with the continuity 
equation\footnote{
  Strictly speaking, neglected in Eq.~(\ref{eq:SDE}) is
  the coupling of the soft mode with the momentum-density modes,
  which plays an important role in describing the critical dynamics
  more precisely \cite{Minami:2011un}.
  This effect is discussed in Ref.~\cite{Kapusta:2012zb}.
}.
As we will see in the next section, the time evolution of the 
critical mode then depends on their length scale~$\Delta y$.
This fact makes the problem complicated because the critical mode can 
no longer be regarded as a spatially uniform mode.
At the same time, however, this diffusion property 
opens a possibility to study the critical fluctuation through
$\Delta y$
dependence
in experiments, as we will see later.

We also remark that our model can describe the fluctuations 
throughout the time evolution of the hot medium including the critical 
region as well as late stages.
This property is advantageous in understanding dynamics behind 
experimental observables.

\section{Analytic properties}
\label{sec:analytic}

In this section, we formally solve the 
SDE~(\ref{eq:SDE}) and study general properties of the 
second-order cumulant and correlation function analytically.

\subsection{Solution of SDE}
\label{subsec:formalism}

Defining the Fourier transform of $n(y,\tau)$ via
$n(q,\tau)=\int dy \ e^{-iqy} n(y,\tau)$,
the formal solution of Eq.~(\ref{eq:SDE}) with the initial condition
$n(q,\tau_0) $ at $\tau=\tau_0$ is obtained as 
\begin{eqnarray}
\label{eq:sol}
n(q,\tau)&=&n(q,\tau_0) e^{-q^2[d(\tau_0,\tau)]^2/2}\nonumber\\
& &+\int_{\tau_0}^{\tau}d\tau^{\prime} \  iq\xi(q,\tau^{\prime})
e^{-q^2[d(\tau^{\prime},\tau)]^2/2},
\end{eqnarray}
where
\begin{align}
d(\tau_1,\tau_2)
= \Big[2\int_{\tau_1}^{\tau_2}d\tau^{\prime} 
D(\tau^{\prime})\Big]^{1/2}
\label{eq:d}
\end{align}
denotes the diffusion ``length'' in
rapidity space from $\tau_1$ to $\tau_2$ with $\tau_1\le\tau_2$.
The diffusion length $d(\tau_1,\tau_2)$ is a monotonically increasing 
(decreasing) function of $\tau_2$ ($\tau_1$),
satisfying the boundary condition $d(\tau,\tau)=0$.
The correlation function at proper time $\tau$ is obtained by 
taking the average of the product of Eq.~(\ref{eq:sol}) as\footnote{
This procedure to solve the stochastic equation corresponds to 
Stratonovich integral \cite{Gardiner}.
There is alternative method called Ito stochastic integral.
These two stochastic integrals give the same result
for Eq.~(\ref{eq:SDE}).}
\begin{align}
\label{eq:cor-fourier1}
\lefteqn{ \langle \delta n(q_1,\tau) \delta n(q_2,\tau)\rangle}
\quad &
\nonumber \\
=&\langle \delta n(q_1,\tau_0)\delta n(q_2,\tau_0)\rangle
e^{-(q_1^2+q_2^2)[d(\tau_0,\tau)]^2/2}
\nonumber\\
&+\int_{\tau_0}^{\tau}d\tau_1 d\tau_2 
\langle i q_1 \xi(q_1,\tau_1) i q_2 \xi(q_2,\tau_2) \rangle 
\nonumber \\
&
\times e^{-q_1^2[d(\tau_1,\tau)]^2/2}\, e^{-q_2^2[d(\tau_2,\tau)]^2/2},
\end{align}
where we have used
$\langle n(y_1,\tau_0)\xi(y_2,\tau)\rangle=0$
for $\tau_0\le\tau$.

To proceed further, 
we assume that the initial fluctuation
satisfies the locality condition,
\begin{align}
\langle \delta n(y_1,\tau_0) \delta n(y_2,\tau_0) \rangle
= \chi(\tau_0) \delta(y_1-y_2).
\label{eq:locality}
\end{align}
Indeed, this condition should hold in a thermal medium 
at the length scale at which the extensive property of
thermodynamic functions is satisfied \cite{Asakawa:2015ybt}.
We then obtain $\langle \delta n(q_1,\tau_0)\delta n(q_2,\tau_0)\rangle
=2\pi\delta(q_1+q_2)\chi(\tau_0)$,
and Eq.~(\ref{eq:cor-fourier1}) is calculated to be
\begin{align}
\label{eq:cor-fourier2}
\lefteqn{ \langle \delta n(q_1,\tau) \delta n(q_2,\tau)\rangle}
\quad &
\nonumber \\
=
& 2\pi \delta(q_1+q_2)
\bigg( \chi (\tau_0) \, e^{-q_1^2 [d(\tau_0,\tau)]^2} 
\nonumber \\
&
+ 2 q_1^2\int_{\tau_0}^{\tau}d\tau^{\prime} \chi(\tau^{\prime})
D(\tau^{\prime})e^{-q_1^2[d(\tau^{\prime},\tau)]^2} \bigg).
\end{align}
The correlation function in $y$ space
is obtained from Eq.~(\ref{eq:cor-fourier2})
as 
\begin{align}
\lefteqn{ \langle \delta n(y_1,\tau) \delta n(y_2,\tau)\rangle }
\quad&
\nonumber \\
=& \chi(\tau_0) G(y_1-y_2;2 d(\tau_0,\tau))
\nonumber\\
&+\int_{\tau_0}^{\tau} d\tau^{\prime} \chi(\tau^{\prime})
\frac{d}{d\tau'} G\big(y_1-y_2;2d(\tau^{\prime},\tau)\big)
\label{eq:cor}
\\
=&\chi(\tau)\delta(y_1-y_2) \nonumber\\
&-\int_{\tau_0}^{\tau}d\tau^{\prime} \chi^{\prime}(\tau^{\prime})
G\big(y_1-y_2;2d(\tau^{\prime},\tau)\big),
\label{eq:cor'}
\end{align}
where $\chi'(\tau) =d \chi(\tau) /d\tau$
and we have defined the normalized Gauss distribution 
\begin{eqnarray}
\label{eq:I}
G(\bar{y};d) = \frac1{\sqrt{\pi} d} e^{-\bar{y}^2/d^2}.
\end{eqnarray}
We note that the correlation function
depends on $D(\tau)$ only through the diffusion length $d(\tau',\tau)$.

\begin{figure}
\begin{center}
	\includegraphics[keepaspectratio, angle=-90, clip,  width=\linewidth]{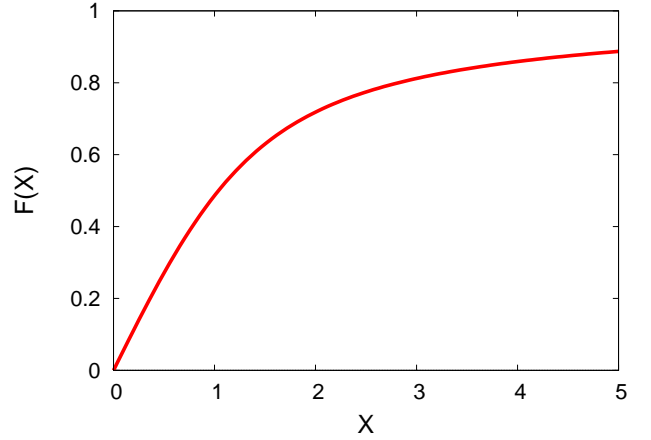}
	\caption{Function $F(X)$ defined in Eq.~(\ref{eq:F}).}
	\label{fig:F(X)}
\end{center}
\end{figure}

By substituting Eq.~(\ref{eq:cor'}) into Eq.~(\ref{eq:<Q^2>cor}),
the second-order cumulant is calculated
to be
\begin{align}
\label{eq:cum}
\frac{\langle Q_{\Delta y}(\tau)^2\rangle_{\rm c}}{\Delta y}
=\chi(\tau) 
- \int_{\tau_0}^{\tau}d\tau' \chi^{\prime}(\tau')
F\Big( \frac{\Delta y}{2d(\tau',\tau) } \Big),
\end{align}
where 
\begin{align}
\label{eq:F}
F(X)
&= \frac2{\sqrt{\pi}} \int_{0}^X dz \Big(1-\frac{z}X \Big)e^{-z^2}
\nonumber \\
&={\rm erf}(X)+\frac{e^{-X^2}-1}{\sqrt{\pi}X}
\end{align}
with the error function
${\rm erf}(x)= \tfrac{2}{\sqrt{\pi}}\int_0^x dz e^{-z^2}$.
The behavior of $F(X)$ is shown in Fig.~\ref{fig:F(X)}.
As is shown in the figure, $F(X)$ is a monotonically increasing function 
satisfying 
\begin{align}
\lim_{X\to 0}F(X)=0, \quad \lim_{X\to\infty}F(X)=1.
\label{eq:Flimit}
\end{align}

\subsection{Properties of fluctuation observables}
\label{subsec:property}

From Eqs.~(\ref{eq:cum}) and (\ref{eq:cor'}), we find 
several notable features in the rapidity dependences
of the cumulant and the correlation function.

First, we consider the behavior of 
$\langle Q_{\Delta y}(\tau)^2 \rangle_{\rm c}$
in the small and large $\Delta y$ limits. 
Using Eq.~(\ref{eq:Flimit}), the cumulant in these limits is easily 
calculated to be
\begin{align}
  \frac{\langle Q_{\Delta y}(\tau)^2 \rangle_{\rm c}}{\Delta y}
  \xrightarrow[\Delta y\to0]{} &
  \chi(\tau),
  \label{eq:Q:Dy->0}
  \\
  \frac{\langle   Q_{\Delta y}(\tau)^2\rangle_{\rm c}}{\Delta y}
  \xrightarrow[\Delta y\to\infty]{} &
  \chi(\tau) 
  - \int_{\tau_0}^{\tau}d\tau' \chi^{\prime}(\tau') 
  \nonumber \\
  =& \chi(\tau_0).
  \label{eq:Q:Dy->inf}
\end{align}
These results show that $\langle Q_{\Delta y}(\tau)^2 \rangle_{\rm c}/\Delta y$ 
takes the local-equilibrium 
value $\chi(\tau)$ in the small $\Delta y$ limit,
while it recovers the initial value in the opposite limit.
This shows that the relaxation toward the equilibrium state
is sufficiently fast 
as $\Delta y$ becomes smaller,
but it becomes arbitrarily slow 
with increasing $\Delta y$.
The latter means that equilibration of the conserved-charge 
fluctuation over the whole system cannot be achieved within a finite time,
because it takes infinite time to transport a charge from one end
to the other.
The analysis here also implies that 
$\langle Q_{\Delta y}(\tau)^2 \rangle_{\rm c}/\Delta y$ with smaller (larger)
$\Delta y$ bears the information of
$\chi(\tau)$ at later (earlier) $\tau$.
This suggests that one can study the $\tau$ dependence of $\chi(\tau)$ 
from the $\Delta y$ dependence of 
$\langle Q_{\Delta y}(\tau)^2 \rangle_{\rm c}/\Delta y$
\cite{Kitazawa:2013bta}.

Second, 
when $\chi(\tau)$ increases (decreases) monotonically in $\tau$,
$\chi'(\tau)\ge0$ ($\le 0$), then
$\langle Q_{\Delta y}(\tau)^2 \rangle_{\rm c}/\Delta y$  for a given $\tau$
is a monotonically 
decreasing (increasing) function of $\Delta y$:
\begin{align}
  \chi'(\tau)
  \left \{ \begin{array}{cc} \ge 0 \\ \le 0 \end{array}  \right .
\quad \Rightarrow \quad
\frac{d~}{d \Delta y} 
\frac{\langle Q_{\Delta y}(\tau)^2\rangle_{\rm c}}{\Delta y}
\left \{ \begin{array}{cc} \le 0 \\ \ge 0 \end{array} \right .
.
\label{eq:dQ<0}
\end{align}
This can be easily shown from Eq.~(\ref{eq:cum}) and the fact  that $F(X)$ is a monotonically increasing function.
Taking the contraposition of Eq.~(\ref{eq:dQ<0}),
one concludes that $\chi(\tau)$ must 
have at least one extremum when
$\langle Q_{\Delta y}(\tau)^2 \rangle_{\rm c}/\Delta y$ 
is non-monotonic as a function of $\Delta y$. 
In particular, 
\begin{align}
\boxed{
\begin{array}{c}
\langle Q_{\Delta y}(\tau)^2 \rangle_{\rm c}/\Delta y
\\
\mbox{has a local maximum} \\
\mbox{as a function of $\Delta y$}
\end{array}
}
\Rightarrow
\boxed{
\begin{array}{c}
\chi(\tau)
\\
\mbox{has a local maximum} \\
\mbox{as a function of $\tau$}
\label{eq:nonmono-cum}
\end{array}
}
\end{align}
The same argument also applies
to the correlation function.
From the fact that $G(\bar{y},d)$ monotonically decreases
as $\bar{y}$ increases, it is again easy to show that 
\begin{align}
  \chi'(\tau)
  \left \{ \begin{array}{cc} \ge 0 \\ \le 0 \end{array}  \right .
\quad \Rightarrow \quad
\frac{d~}{d \bar{y}} 
\langle \delta n(\bar{y},\tau) \delta n(0,\tau)\rangle
\left \{ \begin{array}{cc} \ge 0 \\ \le 0 \end{array}  \right .
,
\label{eq:dn<0}
\end{align}
for $\bar{y}>0$.
From the contraposition of Eq.~(\ref{eq:dn<0}), one obtains
\begin{align}
\boxed{
\begin{array}{c}
\langle \delta n(\bar{y},\tau) \delta n(0,\tau)\rangle
\\
\mbox{has a local minimum} \\
\mbox{as a function of $\bar{y}$}
\end{array}
}
\Rightarrow
\boxed{
\begin{array}{c}
\chi(\tau)
\\
\mbox{has a local maximum} \\
\mbox{as a function of $\tau$}
\label{eq:nonmono-cor}
\end{array}
}
\end{align}

The properties (\ref{eq:nonmono-cum}) and (\ref{eq:nonmono-cor}) 
are quite useful in extracting
the $\tau$ dependence of $\chi(\tau)$
in relativistic heavy ion collisions.
If the experimental results of 
$\langle Q_{\Delta y}(\tau)^2 \rangle/\Delta y$ and/or
$\langle \delta n(\bar{y},\tau) \delta n(0,\tau)\rangle$ show 
non-monotonic behavior as a function of rapidity, 
this immediately confirms the existence of  non-monotonicity
in $\chi(\tau)$ as a function of $\tau$.
It is known that the susceptibilities of baryon number and electric charge
have a peak structure along the phase boundary 
around the QCD critical point \cite{Hatta:2002sj,Asakawa:2009aj}.
The peak in $\langle Q_{\Delta y}(\tau)^2 \rangle_{\rm c}/\Delta y$ 
or $\langle \delta n(\bar{y},\tau) \delta n(0,\tau)\rangle$ 
serves as an experimental signal for this critical enhancement.
This is the most important conclusion of this paper.

It should be kept in mind that the inverses of Eqs.~(\ref{eq:nonmono-cum})
and (\ref{eq:nonmono-cor}) do not necessarily hold.
That is, even if $\chi(\tau)$ is a non-monotonic function of $\tau$,
there is a possibility that 
$\langle Q_{\Delta y}(\tau)^2 \rangle_{\rm c}/\Delta y$ 
and $\langle \delta n(\bar{y},\tau) \delta n(0,\tau)\rangle$ 
are monotonic.
Therefore, from the monotonic behavior of
$\langle Q_{\Delta y}(\tau)^2 \rangle_{\rm c}/\Delta y$ 
and/or $\langle \delta n(\bar{y},\tau) \delta n(0,\tau)\rangle$
one cannot conclude anything about the $\tau$
dependence of~$\chi(\tau)$.
In Appendix~\ref{sec:discussion}, 
we discuss the condition for 
the appearance of the non-monotonic behaviors in these functions
in more detail.

\subsection{Comment on higher order cumulant}

Using SDE (\ref{eq:SDE}), 
it is possible to calculate the time evolution of third and still
higher order cumulants and correlation functions.
As is easily shown, however, the higher order correlation functions
$\langle \delta n(y_1,\tau) \delta n(y_2,\tau) \cdots 
\delta n(y_N,\tau)\rangle$, and accordingly the higher order cumulants
$\langle Q_{\Delta y}(\tau)^N \rangle_{\rm c}$, too, 
vanish in the $\tau\to\infty$ limit for $N\ge3$ \cite{Asakawa:2015ybt};
the fluctuations of $n(y,\tau)$ in equilibrium 
described by Eq.~(\ref{eq:SDE}) obey the Gaussian distribution.

Because of this property, 
SDE~(\ref{eq:SDE}) is not capable of describing 
the relaxation of 
higher order fluctuations toward nonzero non-Gaussian equilibrium values.
In relativistic heavy ion collisions, observed higher order cumulants
take values close to their nonzero equilibrium values 
\cite{Asakawa:2015ybt}.
If this result is a consequence of the relaxation process in the
hadronic medium, the relaxation cannot be described by 
the SDE~(\ref{eq:SDE}).
This is one of the reasons why we limit our attention to the 
second-order cumulant and correlation function in the present study,
in spite of the useful properties of the higher order cumulants 
\cite{Ejiri:2005wq,Stephanov:2008qz,Asakawa:2009aj,Friman:2011pf}.
To describe the relaxation of higher-order cumulants toward nonzero 
non-Gaussianity, different approaches are needed.
In Ref.~\cite{Kitazawa:2013bta}, for example, the non-interacting 
Brownian particle model is employed to describe this process.

\section{Model of collision evolution}
\label{sec:time}

In the previous section we showed that non-monotonic behavior of 
$\langle Q_{\Delta y}(\tau)^2 \rangle_{\rm c}/\Delta y$ and/or
$\langle \delta n(\bar{y},\tau) \delta n(0,\tau)\rangle$, 
if observed, is a direct experimental evidence for
the existence of a peak structure in susceptibility $\chi(\tau)$.
In the rest of this paper, we demonstrate the appearance
of the non-monotonic behavior by studying the behavior of 
$\langle Q_{\Delta y}(\tau)^2 \rangle_{\rm c}/\Delta y$ and
$\langle \delta n(\bar{y},\tau) \delta n(0,\tau)\rangle$ 
with a phenomenological parametrization of 
$\chi(\tau)$ and $D(\tau)$ for a collision event evolution 
passing near and away from the QCD critical point.
In this section we first introduce the model for $\chi(\tau)$ and $D(\tau)$.
Then, the time evolution of fluctuation is studied in the next section.

In this study,
we write the susceptibility and the diffusion coefficient
at temperature $T$
as a sum of their singular and regular contributions:
\begin{align}
\label{eq:sus}
\chi(T)
&= \chi^{\rm cr}(T)+\chi^{\rm reg}(T),
\\
\label{eq:D}
\frac1{D(T)}
&= \tau^2\Big(
\frac1{D_{\rm C}^{\rm cr}(T)}+ \frac1{D_{\rm C}^{\rm reg}(T)} \Big),
\end{align}
where $\chi^{\rm cr}(T)$ and $\chi^{\rm reg}(T)$ denote the singular and
regular parts of susceptibility per unit rapidity, respectively.
We also define the singular and regular parts of diffusion coefficients 
$D_{\rm C}^{\rm cr}(T)$ and $D_{\rm C}^{\rm reg}(T)$ in Cartesian coordinate.
We then parametrize the map of the evolution time to the temperature
$T=T(\tau)$ to fix the $\tau$ dependences.

\subsection{Singular part}
\label{subsec:sin}

First, we discuss the singular parts 
$\chi^{\rm cr}(T)$ and $D^{\rm cr}(T)$.
It is known that the QCD critical point
belongs to the same static universality class
as the 3D Ising model.
The magnetization $M$ of the Ising model
as a function of the reduced temperature $r$ and the dimensionless
magnetic field $H$ 
near the critical point is parametrized with
the two variables $R\geq0$ and $\theta$
in the linear parametric model~\cite{Guida:1996ep,Schofield:1969zza} as 
\begin{eqnarray}
M(R,\theta)&=&m_0R^{\beta}\theta,
\label{eq:M}
\end{eqnarray}
where $r$ and $H$ are expressed as
\begin{eqnarray}
r(R,\theta)&=&R(1-\theta^2),\\
H(R,\theta)&=&h_0R^{\beta\delta}h(\theta)=h_0R^{\beta\delta}\theta(3-2\theta^2).
\end{eqnarray}
The critical point is located at $r=H=0$.
The crossover $(r>0,H=0)$ and first-order transition $(r<0,H=0)$ lines
correspond to $\theta=0$ and $|\theta|=\sqrt{3/2}$ with $R>0$, 
respectively.
We adopt approximate values $\beta=1/3$ and $\delta=5$
for the Ising critical exponents \cite{Guida:1996ep}.
From Eq.~(\ref{eq:M}), one can calculate the 
magnetic susceptibility as 
\begin{eqnarray}
 \label{eq:sus1}
 \chi_M(r,H) =
\frac{\partial M(r,H)}{\partial H}\bigg|_{r}
 \label{eq:sus2}
 = \frac{m_0}{h_0}\frac1{R^{4/3}(3+2\theta^2)}.
\end{eqnarray}
As the susceptibility of a 
conserved charge $\chi$ near the QCD critical point should share 
the same critical behavior as $\chi_M(r,H)$, 
we set \cite{Stephanov:2011pb,Mukherjee:2015swa,Bzdak:2016sxg} 
\begin{align}
\frac{\chi^{\rm cr}(r,H)}{\chi^{\rm H}} 
= c_{\rm c} \chi_M(r,H)
= c_{\rm c} \frac{m_0}{h_0}\frac1{R^{4/3}(3+2\theta^2)},
\label{eq:chi^cr}
\end{align}
with a dimensionless proportionality constant $c_{\rm c}$.
The susceptibility in the hadronic medium $\chi^{\rm H}$ will be
defined in Sec.~\ref{subsec:sreg}.
We fix the normalization constants $m_0$ and $h_0$ 
by imposing $M(r=-1, H=0^+)=1$ and $M(r=0, H=1)=1$. 

In reality, the finite system size effect in heavy ion collisions 
prevents the divergence of $\chi$~\cite{Stephanov:1999zu}. 
Nevertheless we ignore this effect because
the growth of fluctuation would be limited more severely by
the finiteness of the evolution time owing to the critical 
slowing down \cite{Berdnikov:1999ph}.

For determining $D_{\rm C}^{\rm cr}$,
we employ the dynamic universality argument
\cite{Hohenberg:1977ym}.
Since the QCD critical point belongs to the model H \cite{Son:2004iv}
in the classification of Ref.~\cite{Hohenberg:1977ym},
the singular part $D_{\rm C}^{\rm cr}$
scales with the correlation length $\xi$ as 
$D_{\rm C}^{\rm cr}\sim\xi^{-2+\chi_\eta+\chi_\lambda}$, where 
the exponents $\chi_\eta$ and $\chi_\lambda$ for model H are 
obtained by the renormalization group calculation
as $\chi_\eta\simeq0.04$ and 
$\chi_\lambda\simeq 0.916$ \cite{Hohenberg:1977ym}.
As the correlation length $\xi$ and the susceptibility $\chi^{\rm cr}$
are related as $\chi^{\rm cr}\sim\xi^{2-\chi_\eta}$,
the singular part $D_{\rm C}^{\rm cr}$ can be expressed
in terms of $\chi^{\rm cr}$:
\begin{eqnarray}
  D_{\rm C}^{\rm cr}(r,H)=d_{\rm c}
  \left[\frac{\chi^{\rm cr}(r,H)}{\chi^{\rm H}}\right]
  ^{(-2+\chi_\eta+\chi_\lambda)/(2-\chi_\eta)}
  \label{eq:D^cr}
\end{eqnarray}
with a proportionality constant $d_{\rm c}$ 
having the dimension of diffusion coefficient.
We set $d_{\rm c}=1$~fm in what follows.
Note that $D_{\rm C}^{\rm cr}$ vanishes at the critical point,
reflecting the critical slowing down.

In our model, we leave the strengths of the critical component
$c_{\rm c}$ as a free parameter.
The reduced temperature $r$ controls the distance of the trajectory
from the critical point.
We will vary $r$ to simulate
the change of the collision energy in the next section.

\subsection{Parametrizing the medium evolution}

To utilize the above universality argument
for describing heavy ion collision events,
we need a map between the Ising variables $(r, H)$ and the physical variables
$(T,\mu)$ in QCD, in addition to a map from the proper time $\tau$
to $(T,\mu)$ at a given collision energy.
To skip these mappings,
we follow a simple approach adopted in 
Refs.~\cite{Berdnikov:1999ph,Mukherjee:2015swa}:
We assume that $(T,\mu)$ in QCD are linearly mapped to the Ising 
variables $(r, H)$ around the critical point,
and that only $H$ changes while $r$ is fixed
during the time evolution.
We write the linear relation between $H$ and $T$ as
\begin{eqnarray}
\label{eq:region}
\frac{T-T_{\rm c}}{\Delta T}=\frac{H}{\Delta H},
\end{eqnarray}
with $T_{\rm c}$ being the critical temperature. 
The ratio $\Delta T/\Delta H$ controls the width of the critical 
region in the QCD phase diagram.
To relate $T$ and $\tau$, 
we assume the one-dimensional Bjorken expansion and 
 conservation of total entropy.
The relation is then obtained as \cite{Mukherjee:2015swa}
\begin{eqnarray}
\label{eq:T-t}
T(\tau)=T_0\left(\frac{\tau_0}{\tau}\right)^{c_{\rm s}^2},
\end{eqnarray}
where $c_{\rm s}^2$ is the sound velocity
and $T_0$ is the initial temperature of
the system at the initial proper time $\tau_0$.
We ignore possible entropy production in the critical region 
in this work.

In our calculation we set the initial temperature $T_0=220$ MeV
at proper time $\tau_0=1.0$~fm, 
the critical temperature of the QCD critical point 
$T_{\rm c}=160$ MeV, and 
the kinetic freeze-out temperature $T_{\rm f}=100$ MeV \cite{Kumar:2014tca},
where we stop the evolution.
The parameters for the critical region in Eq.~(\ref{eq:region}) are set
to $\Delta T/\Delta H=10$~MeV.
For the sound velocity $c_{\rm s}^2$,
we adopt $c_{\rm s}^2 = 0.15$, which is indicated in a lattice calculation 
in the transition region at $\mu=0$~\cite{Bazavov:2014pvz}.
In Eq. (\ref{eq:T-t}), the effect of transverse expansion is not
taken into account. Transverse expansion makes the duration of the
hadron phase shorter. Thus, the following calculation is likely
to estimate the effect of diffusion stronger than the actual one.
As we will see, the suppression of the diffusion will be advantageous
to detecting the critical point in experiments.

\subsection{Regular $+$ singular}
\label{subsec:sreg}

We assume that the susceptibility per unit rapidity $\chi(T)$ 
approaches a constant value
$\chi^{\rm Q}$ ($\chi^{\rm H}$) at high (low) temperature, which
we call quark-gluon plasma (hadronic) value.
We note that the value of $\chi^{\rm Q}$ 
depends on the trajectory in the $(T,\mu)$ plane 
as well as the thermal property \cite{Asakawa:2000wh,Jeon:2000wg};
see also Refs.~\cite{Koch:2008ia,Asakawa:2015ybt}.
We also note that the susceptibility per unit rapidity approaches 
a constant value in the late stage in heavy ion collisions 
because the particle abundances are fixed after the chemical freeze-out. 
In the present study, we use the value 
estimated in Ref.~\cite{Asakawa:2000wh} assuming entropy 
conservation,
\begin{align}
\frac{\chi^{\rm Q}}{\chi^{\rm H}}\simeq0.5.
\label{eq:chi^Q}
\end{align}

For the diffusion coefficient,
we assume that the coefficient in the Cartesian coordinates approaches 
constant values, $D_{\rm C}^{\rm Q}$ and $D_{\rm C}^{\rm H}$, at 
high and low temperatures, respectively.
We take $D_{\rm C}^{\rm Q}=2.0$~fm from an estimate 
in the lattice QCD calculation~\cite{Aarts:2014nba} and 
$D_{\rm C}^{\rm H}=0.6$ fm from 
Ref.~\cite{Sakaida:2014pya}.

The regular parts $\chi^{\rm reg}(T)$ and 
$D_{\rm C}^{\rm reg}(T)$ are then constructed by 
smoothly interpolating these values at high and low temperatures,
\begin{eqnarray}
\label{eq:regsus}
\chi^{\rm reg}(T)
&=&
\chi_0^{\rm H} + (\chi_0^{\rm Q}-\chi_0^{\rm H}) S(T),
\\
\label{eq:regD}
D_{\rm C}^{\rm reg}(T)&=&
D_{0}^{\rm H} + (D_{0}^{\rm Q}-D_{0}^{\rm H}) S(T),
\end{eqnarray}
with 
\begin{align}
S(T) = \frac12 \left (1 + \tanh\left(\frac{T-T_{\rm c}}{\delta T} \right)\right ).
\end{align}
Here $\chi_0^{\rm Q,H}$ and $D_0^{\rm Q, H}$ are determined so that 
$\chi(T)$ and $D_{\rm C}(T)$ coincide with the presumed values 
$\chi^{\rm Q,H}$ and $D_{\rm C}^{\rm Q, H}$ at $T=T_{0,\rm f}$, respectively.
We set the width of the crossover region, $\delta T=10$~MeV.

\begin{figure}
\begin{center}
  \includegraphics[keepaspectratio, angle=-90, clip,  width=\linewidth]{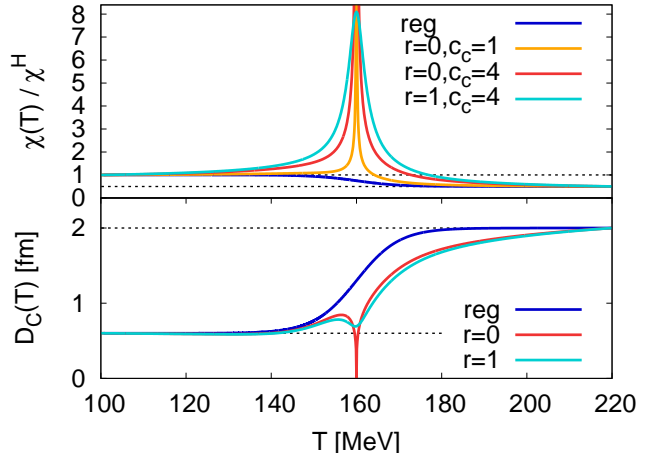}
  \caption{Susceptibility in rapidity space $\chi(T)$ (upper) and diffusion 
    coefficient in the Cartesian coordinate
    $D_{\rm C}$ in units of fm (lower) as a function of $T$ 
    for several values of $r$ and
    $c_{\rm c}$. The regular parts of the susceptibility $\chi^{\rm reg}(T)$ 
    and of the diffusion coefficient $D_{\rm C}^{\rm reg}(T)$, labeled ``reg'',
    are also shown. The dashed lines show the initial and final values.}
  \label{fig:chiD}
\end{center}
\end{figure}

In Fig.~\ref{fig:chiD}, we plot 
the susceptibility $\chi(T)/\chi^{\rm H}$ and 
the diffusion coefficient $D_{\rm C}(T)$ as a function of $T$
for several values of $r$ and $c_{\rm c}$ with $d_{\rm c}=1$.
The upper panel of Fig.~\ref{fig:chiD} shows
that $\chi(T)/\chi^{\rm H}$ for $r=0$ diverges 
at $T=T_{\rm c}$. The sharp peak around $T_{\rm c}$ remains 
even for $r=1$ with $c_{\rm c}=4$.
The regular part Eq.~(\ref{eq:regsus}), labeled  ``reg'', is also shown
for comparison.
The lower panel of Fig.~\ref{fig:chiD} shows
that $D_{\rm C}(T)$ with the singular part vanishes at $T=T_{\rm c}$ for $r=0$, 
which is a manifestation of the critical slowing down.

\section{Effects of criticality on observables}
\label{sec:real} 

Now, we analyze the time evolution of the fluctuation and correlation 
function using the parametrization obtained in the previous section,
and study the effect of the QCD critical point on observables.

One remark concerned with 
the experimentally-observed fluctuations is that the diffusion 
described by the SDE (\ref{eq:SDE}) proceeds in coordinate space,
but the experimental measurements are performed in momentum space.
The imperfect correlation between the two rapidities owing to thermal motion
gives rise to the ``thermal blurring'' effect 
\cite{Ohnishi:2016bdf,Asakawa:2015ybt}.
For nucleons, this effect increases the apparent diffusion
length in rapidity space by about $0.25$ 
at and after the thermal freeze-out~\cite{Ohnishi:2016bdf}.
We take account of this effect in the subsequent analyses.

In this section,
we show the numerical results of the cumulant and correlation 
function in the following normalized forms,
\begin{align}
K(\Delta y) =& 
\frac{ \langle Q_{\Delta y}(\tau)^2 \rangle_{\rm c} }
{ \langle Q_{\Delta y}^2\rangle_{\rm c,H} }
= \frac{ \langle Q_{\Delta y}(\tau)^2 \rangle_{\rm c} }
{ \chi^{\rm H} \Delta y },
\label{eq:K}
\\
C(\bar{y}) =& 
\frac{ \langle \delta n(\bar y,\tau) \delta n(0,\tau)\rangle }{\chi^{\rm H}},
\label{eq:C}
\end{align}
where $\langle Q_{\Delta y}^2\rangle_{\rm c,H}=\chi^{\rm H} \Delta y$
is the cumulant in the equilibrated hadronic medium.

\subsection{Non-critical trajectory}
\label{subsec:reg}

\begin{figure}
  \includegraphics[keepaspectratio, angle=-90, clip, width=\linewidth]{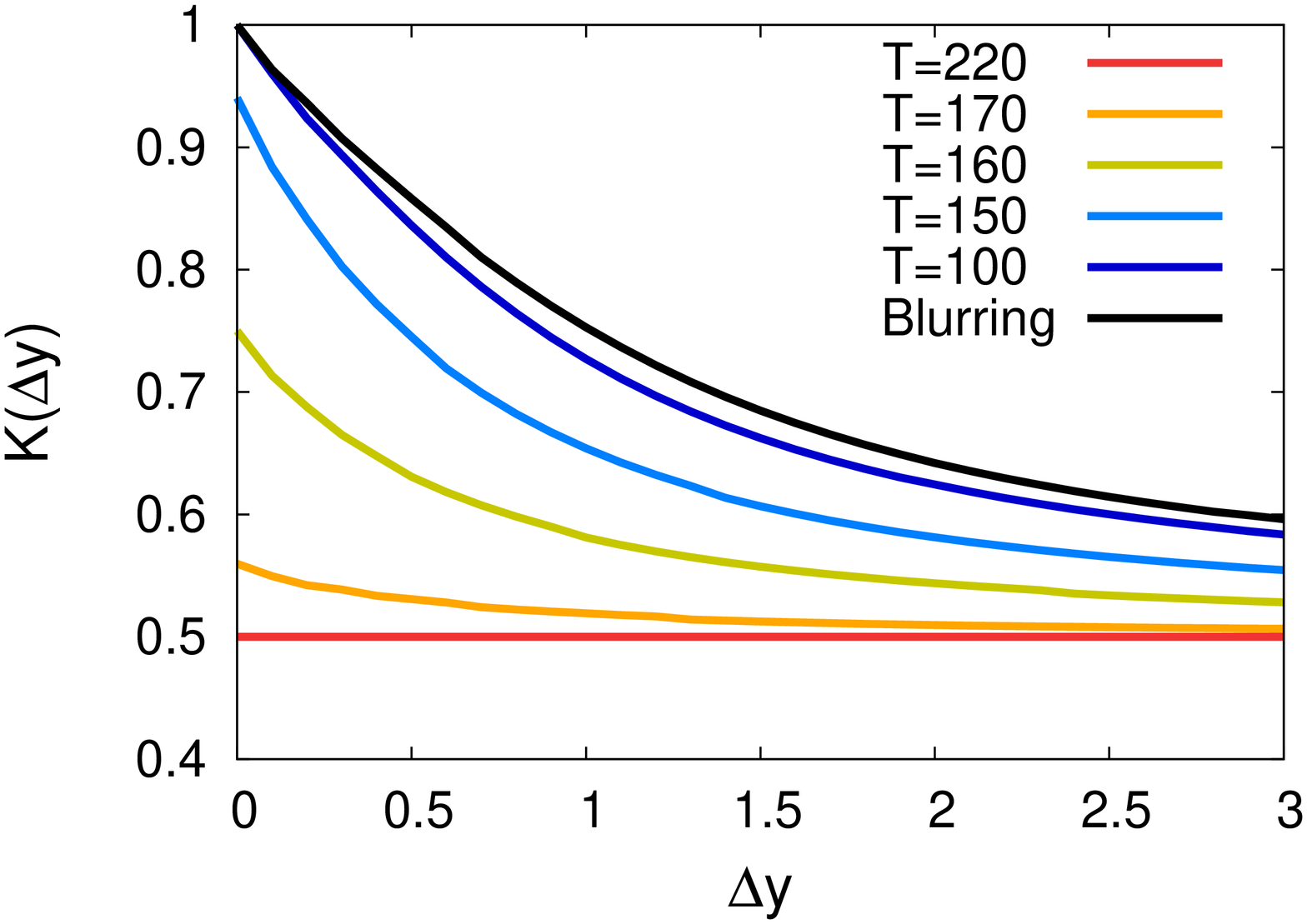}
  \includegraphics[keepaspectratio, angle=-90, clip, width=\linewidth]{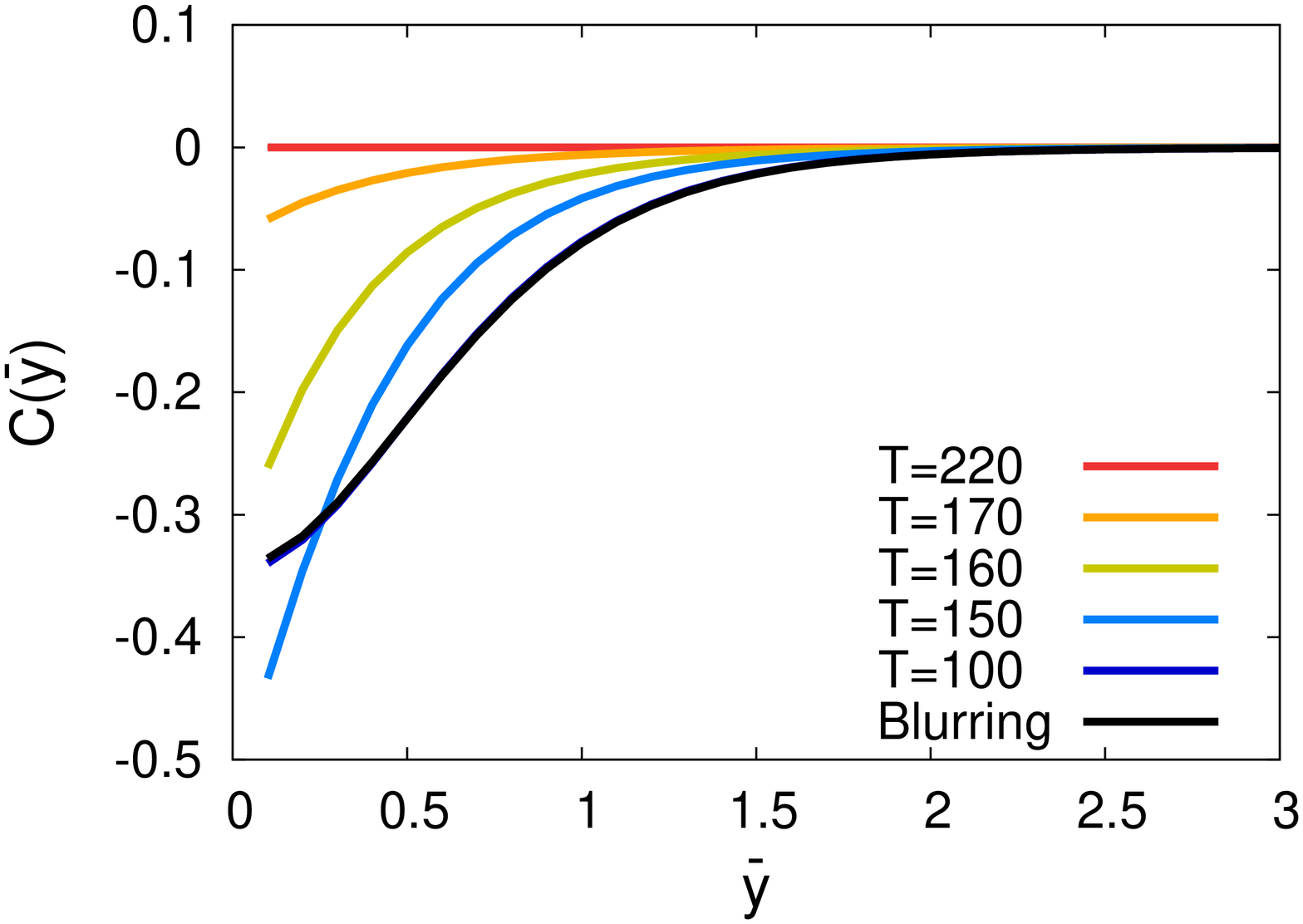}
  \caption{Time slices of second-order cumulant (upper) and correlation 
    function (lower) for a non-critical trajectory.}
  \label{fig:reg}
\end{figure}

First, we study the case without the singular parts by setting
$\chi(\tau)=\chi^{\rm reg}(T(\tau))$ and $D(\tau)=D^{\rm reg}(T(\tau))$.
This corresponds to the collision events which pursue a trajectory 
away from the critical point in the crossover region 
(or, in the QCD phase diagram without a first order phase transition).
As in Fig.~\ref{fig:chiD}, $\chi^{\rm reg}(T)$ behaves monotonically 
as a function of $T$ in this case.

In Fig.~\ref{fig:reg}, we show the results of 
$K(\Delta y)$ and $C(\bar{y})$
for several values of $T$ from the initial temperature $T_0=220$~MeV 
to the kinetic freeze-out $T_{\rm f}=100$~MeV,
together with the result after the thermal blurring.
Note that the result only after thermal blurring
can be compared with experimental results.
The other results are shown to understand the time evolution of these 
quantities.

At the initial time with $T=220$~MeV, $K(\Delta y)$ is given 
by a constant, while $C(\bar{y})$ vanishes, 
in accord with the locality condition Eq.~(\ref{eq:locality}).
As $T$ is lowered, nontrivial structures emerge in these
functions.
As discussed in Sec.~\ref{sec:model}, 
$K(\Delta y)$ at $\Delta y=0$ is equal to its thermal value, i.e. 
$\chi(T)/\chi^{\rm H}$, which increases
monotonically with time in this non-critical case.
$K(\Delta y)$ at nonzero $\Delta y$ 
follows this trend but the increase is slower because
of the finite diffusion time.
As a result, the cumulant of a conserved charge depends on
$\Delta y$ strongly.
We also note that $K(\Delta y)$ decreases monotonically with increasing
$\Delta y$,
which is consistent with the statement, Eq.~(\ref{eq:dQ<0}), which tells us
that $K(\Delta y)$ should be monotonic when $\chi(\tau)$ is monotonic.
$K(\Delta y)$ approaches its initial value $\chi^{\rm Q}/\chi^{\rm H}$
as $\Delta y$ increases.
Notice that the behavior of $K(\Delta y)$ after thermal blurring
is qualitatively consistent with the experimental result 
of the second-order cumulant of net-electric charge observed 
at the Large Hadron Collider \cite{ALICE}.

The $\Delta y$ dependence of the second-order cumulant has been 
studied in Refs.~\cite{Shuryak:2000pd,Kitazawa:2013bta,Sakaida:2014pya}.
The analysis of these studies corresponds to the parameter choice
$\delta T=0$ in Eq.~(\ref{eq:regsus}), i.e. $\chi(\tau)$ jumps 
discontinuously at $T_{\rm c}$.
Since our result is qualitatively unchanged from the previous one,
the nonzero width $\delta T$ seems not crucial for the argument here.

The lower panel of Fig.~\ref{fig:reg} shows $C(\bar{y})$.
One finds that this function also behaves monotonically
as a function of $\bar{y}$, which is consistent with Eq.~(\ref{eq:dn<0}).
We also notice that $C(\bar{y})$ always takes a negative value.
This is directly confirmed by substituting $\chi'(\tau)>0$ 
into Eq.~(\ref{eq:cor'}).

\subsection{Trajectory passing through the critical point}
\label{subsec:on}
\begin{figure}
  \includegraphics[keepaspectratio, angle=-90, clip, width=\linewidth]{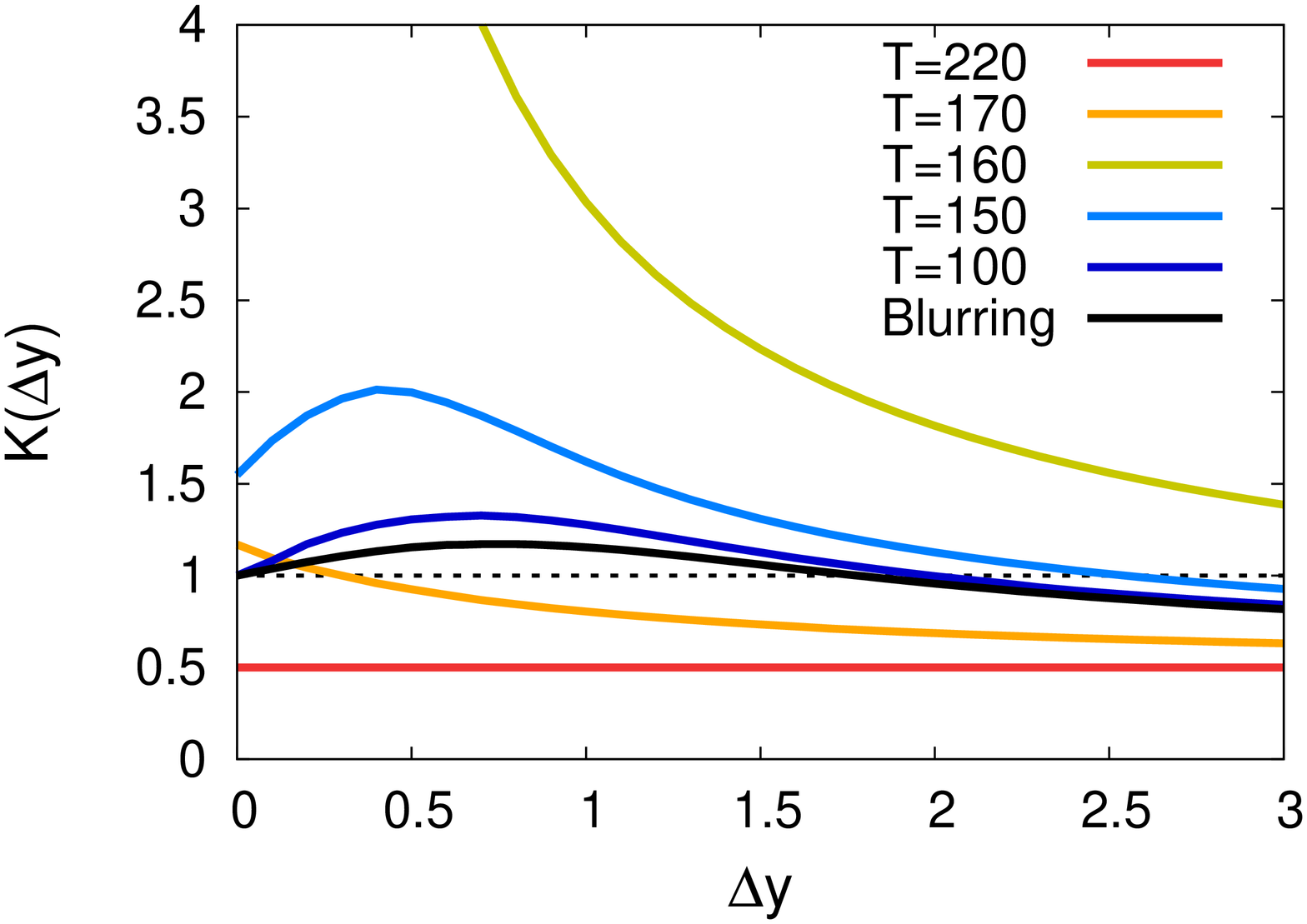}	
  \includegraphics[keepaspectratio, angle=-90, clip, width=\linewidth]{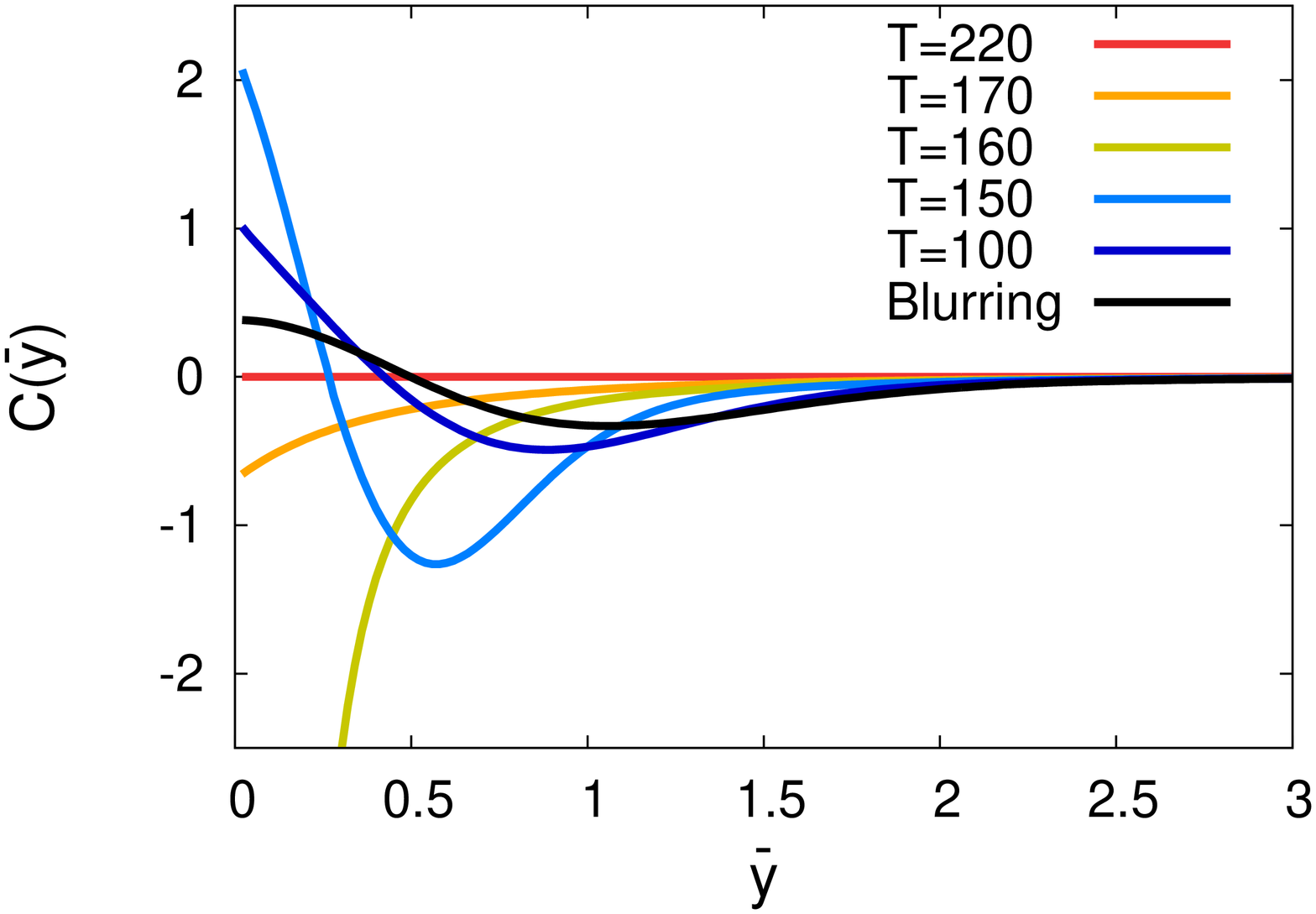}
  \caption{Second-order cumulant (upper) and correlation function (lower) for a trajectory passing through the critical point with $r=0$ and $c_{\rm c}=4$.}
  \label{fig:r=0Cc=4}
\end{figure}

\begin{figure}
  \includegraphics[keepaspectratio, angle=-90, clip, width=\linewidth]{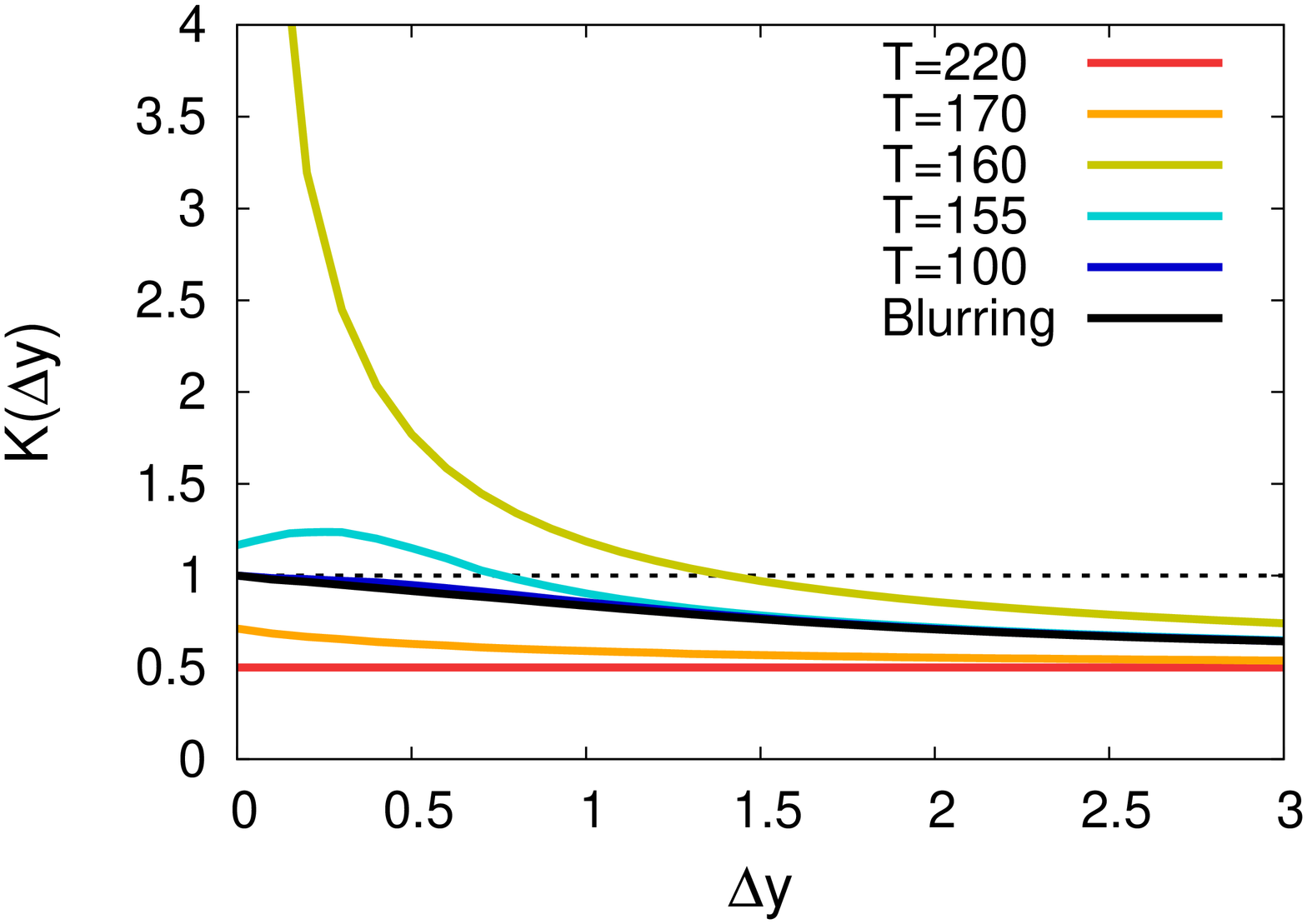}	
  \includegraphics[keepaspectratio, angle=-90, clip, width=\linewidth]{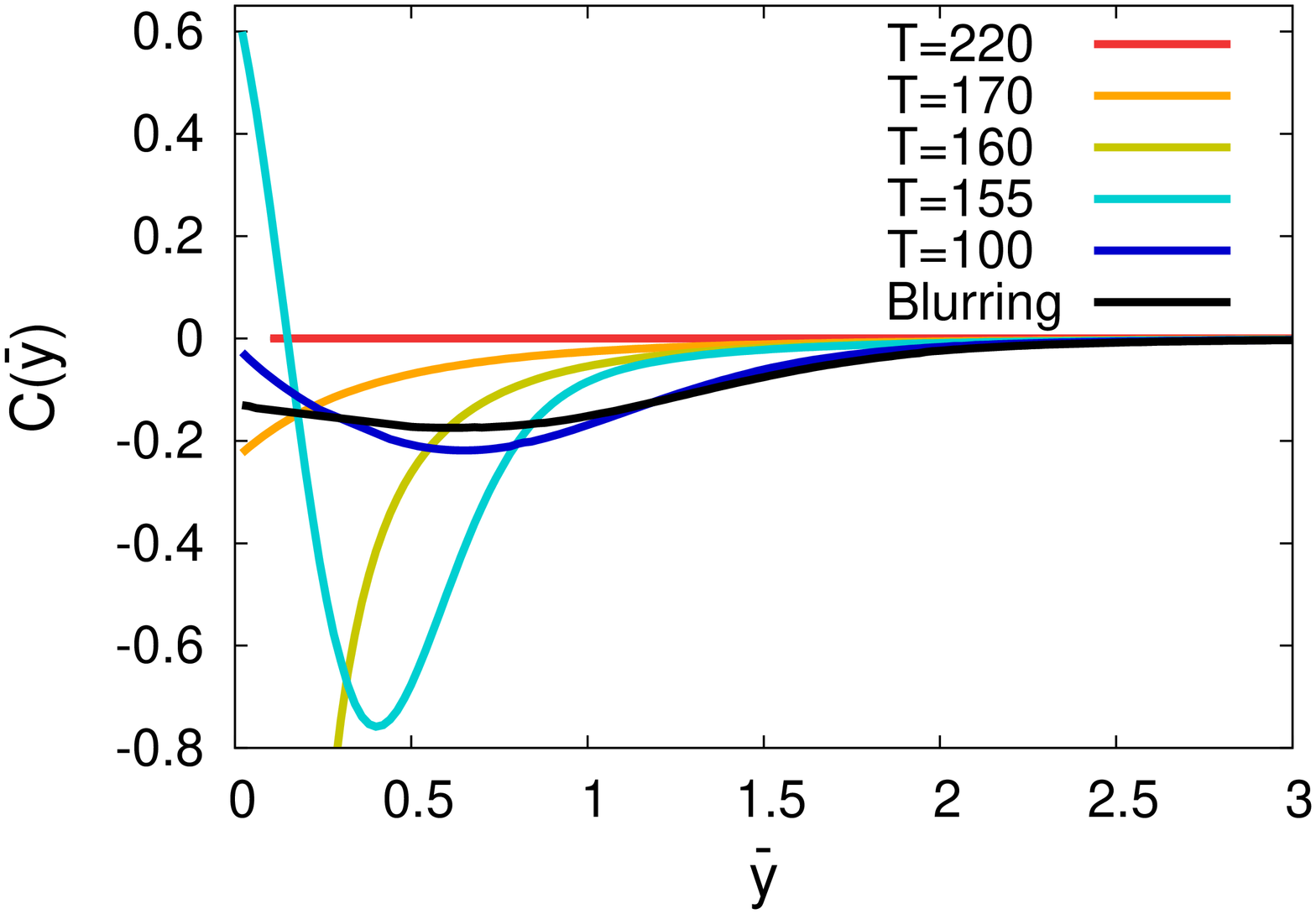}
  \caption{Same as Fig.~\ref{fig:r=0Cc=4} but with weaker criticality 
    with $c_{\rm c}=1$.}
  \label{fig:r=0Cc=1}
\end{figure}

Let us examine the case where the trajectory in heavy ion collisions 
passes right through the critical point ($r=0$).
In Fig.~\ref{fig:r=0Cc=4}, we show the evolution of
$K(\Delta y)$ and $C(\bar{y})$ along the critical
trajectory ($r=0$) for several values of $T$ and 
after thermal blurring with $c_{\rm c}=4$.

In the upper panel of Fig.~\ref{fig:r=0Cc=4},
$K(\Delta y)$ at $T=T_{\rm c}=160$~MeV shows a remarkable enhancement,
which comes from the divergence of $\chi(\tau)$ at $T=T_{\rm c}$
as shown in Fig.~\ref{fig:chiD}.
This figure shows, however, that the cumulant stays finite 
for nonzero $\Delta y$ even at the critical point.
This result is a manifestation of the critical slowing down for 
conserved charges.
We remark that the effect of the critical slowing down is 
dependent on $\Delta y$ as discussed in Sec.~\ref{subsec:property}.
After passing through the critical point,
the value of $K(\Delta y)$ at $\Delta y=0$ decreases rapidly
in accordance with the suppression of $\chi(\tau)$,
while the decrease at nonzero $\Delta y$
is slower because of the slower diffusion for the larger $\Delta y$.
As a consequence, a non-monotonic structure appears
in $K(\Delta y)$.
In Fig.~\ref{fig:r=0Cc=4}, the non-monotonic behavior continues 
to exist in $K(\Delta y)$ until the kinetic freeze-out time and 
even survives the thermal blurring.
Therefore, the non-monotonic behavior of $K(\Delta y)$ can be
observed experimentally in this case.
As discussed in Sec.~\ref{subsec:property},
this non-monotonic behavior, if observed, 
is a direct signal for the existence of the critical enhancement
of $\chi(\tau)$.

An important lesson to learn from this result is that 
the non-monotonic behavior of $K(\Delta y)$ 
can survive whereas the magnitude of fluctuation itself
is almost smeared to the equilibrated hadronic value $K(\Delta y)=1$;
the maximum value of $K(\Delta y)$ after thermal blurring is 
$K(\Delta y)\simeq1.2$ at $\Delta y=0.75$.
This result suggests that the study of the non-monotonicity of
$K(\Delta y)$ is advantageous for the search of the critical enhancement
than the value of $K(\Delta y)$ with fixed $\Delta y$.
Therefore, it is quite an interesting experimental subject to analyze 
its $\Delta y$ dependence. 

From the lower panel of Fig.~\ref{fig:r=0Cc=4}, one can draw 
the same conclusion on the $\bar{y}$ dependence of $C(\bar{y})$:
$C(\bar{y})$ at $\bar{y}\to0$ changes from negative to positive 
around $T_{\rm c}$.
Triggered by this behavior, 
the non-monotonic $\bar{y}$ dependence of $C(\bar{y})$ manifests
itself.
The non-monotonicity again survives thermal blurring, 
suggesting that it can be measured experimentally.

We note that similar non-monotonic behaviors of correlation 
functions are also observed in Ref.~\cite{Kapusta:2012zb} and 
that of the mixed correlation function in Ref.~\cite{Pratt:2012dz}.
The appearance of the non-monotonicity in these studies is understood
completely the same way as that in Sec.~\ref{sec:analytic}.

Next, we consider the case of a weaker critical enhancement by setting
$c_{\rm c}=1$, but still keeping $r=0$.
We show the results in Fig.~\ref{fig:r=0Cc=1}.
Although results above $T=155$~MeV look almost the same as those 
for $c_{\rm c}=4$, the non-monotonicity of $K(\Delta y)$ disappears 
already at $T=100$~MeV.
As the growth of the susceptibility with $c_{\rm c}=1$ is 
weaker, the signal is drowned out 
by the diffusion in the hadronic phase.
This exemplifies that,
as discussed in Sec.~\ref{subsec:property},
the absence of the non-monotonicity in 
$K(\Delta y)$ does not necessarily mean the absence of the peak
structure in $\chi(\tau)$.

The lower panel of Fig.~\ref{fig:r=0Cc=1} shows the
result for the correlation function.
The figure shows that the non-monotonic behavior of $C(\bar{y})$
generated at the critical point survives at $T=100$ MeV and 
even the thermal blurring.
This result suggests that 
the non-monotonic signal in $C(\bar{y})$ is observable 
even when it disappears in $K(\Delta y)$.
In fact, as we will discuss in Appendix.~\ref{sec:discussion}, the 
non-monotonicity in $C(\bar{y})$ is more sustainable 
than that in $K(\Delta y)$.

\subsection{Trajectory passing near the critical point}
\label{subsec:around}

\begin{figure}
  \includegraphics[keepaspectratio, angle=-90, clip, width=\linewidth]{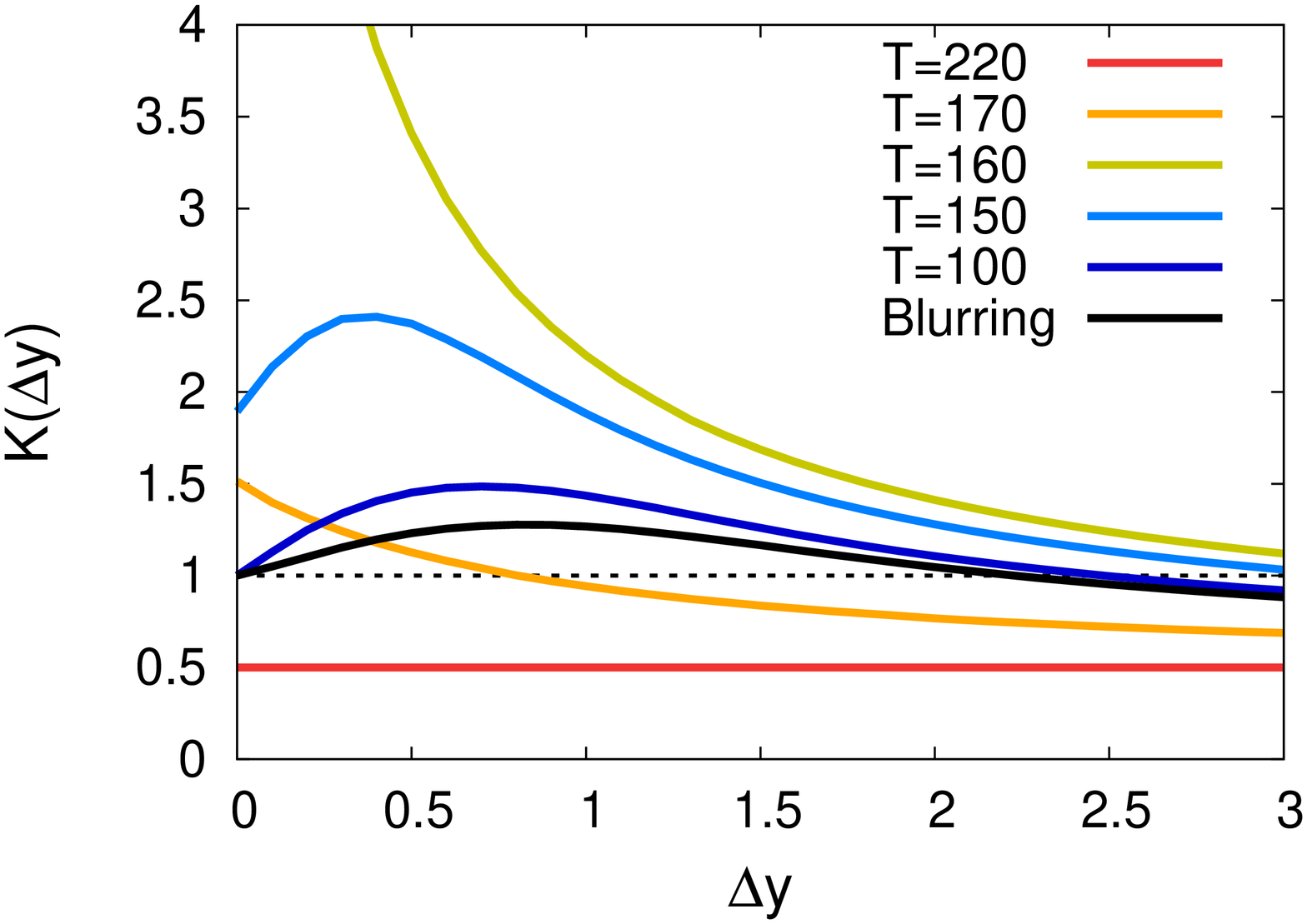}	
  \includegraphics[keepaspectratio, angle=-90, clip, width=\linewidth]{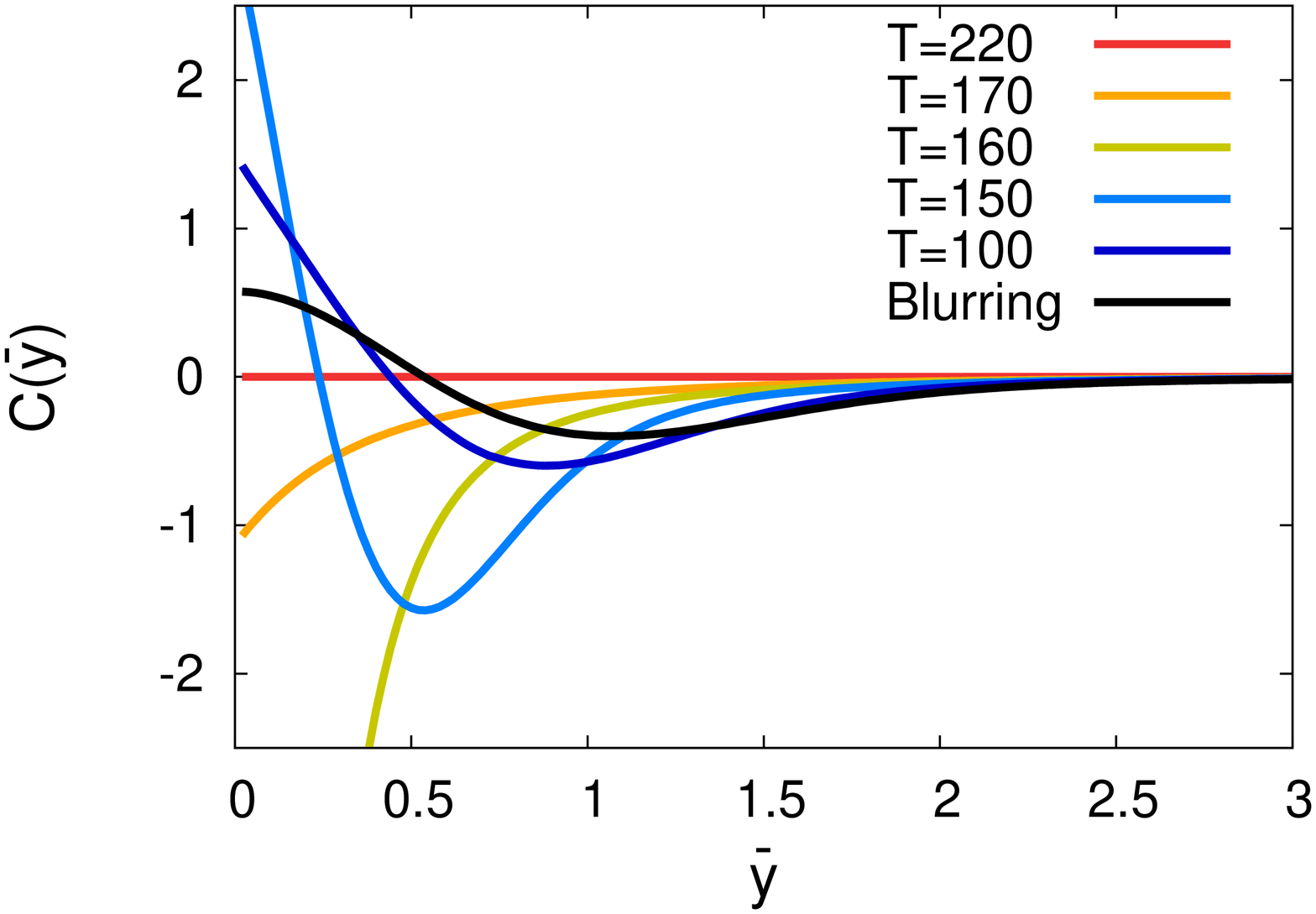}
  \caption{Second-order cumulant (upper) and correlation function (lower) 
    for a trajectory in the crossover region with $r=1$ and $c_{\rm c}=4$.}
  \label{fig:r=1Cc=4}
\end{figure}

Finally, let us study the time evolution of fluctuations for $r>0$, 
which corresponds to the case where the system undergoes a crossover
transition.
Shown in Fig.~\ref{fig:r=1Cc=4} are the results of
$K(\Delta y)$ and $C(\bar{y})$ for $r=1$ and $c_{\rm c}=4$.
The results are qualitatively the same as those in Fig.~\ref{fig:r=0Cc=4}.
By closely comparing these results, 
one finds that the non-monotonic signal with $r=1$ for $T\le150$~MeV
is much clearer than that in Fig.~\ref{fig:r=0Cc=4},
although $\chi(T)$ does not diverge with $r=1$.

There are two reasons behind this result. 
First, $D(\tau)$ for $r=1$ does not vanish because the trajectory 
does not pass right through the critical point (see, Fig.~\ref{fig:chiD}).
Therefore, the critical slowing down for $r=1$ is less important 
than that for $r=0$, 
and the fluctuations can grow faster around $T_{\rm c}$.
Second, $\chi(T)/\chi^{\rm H}$ for $r=1$ and $c_{\rm c}=4$ is 
larger than that for $r=0$ at $T\lesssim 155$~MeV 
in our parametrization, as seen in Fig.~\ref{fig:chiD}.
Therefore, $\chi(T)$ for $r=1$ approaches 
$\chi^{\rm H}$ more slowly.
This behavior makes the non-monotonic peaks in $K(\Delta y)$ and 
$C(\bar{y})$ more prominent.
Note, however, that the second observation may be dependent on 
the parametrization of $\chi(T)/\chi^{\rm H}$.

This argument suggests that the non-monotonic 
signals can be observed even when the trajectory does not pass right
through the critical point.
Moreover, it is possible that the trajectory off the critical point 
is more favorable for the emergence of the non-monotonicity.
However, the signal of the critical enhancement, of course,
weakens and finally disappears
as the trajectory departs further off the critical point. 
In our analysis with $c_{\rm c}=4$, the non-monotonic behavior of 
$K(\Delta y)$ and $C(\bar{y})$ disappears at $r\simeq5$ and $8$,
respectively.

\section{Discussions and a summary}
\label{sec:sum}

The most important conclusion of the present study is 
Eqs.~(\ref{eq:nonmono-cum}) and (\ref{eq:nonmono-cor}), i.e.
the non-monotonic behaviors of $K(\Delta y)$ and/or $C(\bar{y})$, 
if observed, are direct experimental signals of the critical 
enhancement in the susceptibility.
Now, let us consider the application of this conclusion to 
real heavy ion collisions.
Throughout this study we assumed the Bjorken space-time evolution.
This assumption, however, is violated in lower energy collisions.
In particular, in the energy range of the BES program at RHIC,
one has to consider the effect of the violation severely.
For lower energy collisions, the effects of global charge conservation 
\cite{Sakaida:2014pya} arising from the finite system size would also
affect fluctuation observables seriously.
It is expected, however, that these effects do not alter our conclusions
on non-monotonicity as long as the system size in rapidity space is
sufficiently large compared to the diffusion length.
First, the violation of the Bjorken picture makes the correspondence 
between the coordinate and momentum space rapidities worse.
This makes the thermal blurring effect stronger 
\cite{Ohnishi:2016bdf,Asakawa:2015ybt}, and the experimental 
measurement of non-monotonicity would become difficult.
Even in this case, however, the relations Eqs.~(\ref{eq:nonmono-cum}) 
and (\ref{eq:nonmono-cor}) should hold, because the thermal blurring
effect only acts to enhance the diffusion length \cite{Ohnishi:2016bdf}.
From the analysis in Ref.~\cite{Sakaida:2014pya}, it is expected that
the fluctuation around mid-rapidity is not affected by the effect of 
global charge conservation for sufficiently high energy collisions
as long as the system size is sufficiently large.
However, our conclusion may need to
be altered for collisions at BES energies and lower energies such
as those at FAIR, NICA, and J-PARC. 

In this study, we described the critical fluctuation by the stochastic
diffusion equation (\ref{eq:SDE}).
Although this model well describes sufficiently long and slow fluctuations, 
the fluctuations in heavy ion collisions may not be slow enough
compared to the time scale of the medium evolution.
To take account of these effects, the SDE~(\ref{eq:SDE}) has to be modified.
One direction is to include higher order derivative terms.
Another interesting extension is to include the $\sigma$ field 
as a dynamical field, and solve the coupled equation of $n$ and $\sigma$.
Near the critical point, the coupling of $\sigma$ with 
momentum density would also be important \cite{Kapusta:2012zb}.
To deal with these subjects, numerical simulations
of medium evolution will be needed with adopting a certain
model,
for example, the chiral fluid model~\cite{Herold:2016uvv}.

In this study, we investigated the time evolution of the 
second order cumulant and the correlation function of conserved charges
in heavy ion collisions which pass through or near the critical point
focusing on the effects of critical slowing down near the 
critical point and dissipation in the late stages.
We adopted the stochastic diffusion equation 
with critical nature being encoded in the time-dependent 
susceptibility and diffusion coefficient. 
This model can describe the dynamics of the critical mode 
respecting its diffusion property, 
which was not considered in previous studies on
the critical slowing down.
We have pointed out that 
the critical enhancement in susceptibility 
can be observed as the non-monotonic behaviors 
in the second-order cumulant and correlation function.
Our numerical results suggest that these non-monotonic behaviors are 
a more robust experimental signal than the value of these functions themselves.
It is, therefore, quite an interesting experimental subjects to analyze 
the rapidity dependences of these functions in heavy ion collisions.

M.~A. and M.~K. thank V.~Koch and M.~Lisa for inviting them to 
INT workshop ``Exploring the QCD Phase Diagram through Energy Scans'',
Sep.~19 - Oct.~14, 2016, Seattle, USA, and stimulating discussions.
The authors thank M.~Nahrgang and M.~Bluhm for fruitful discussions.
They also thank A.~Bzdak, K.~Redlich, and M.~Stephanov for useful conversations.
This work was supported in part by 
JSPS KAKENHI Grant Numbers 16J01314, 26400272, and 16K05343.

\appendix
\section{Time evolution of the soft mode near the QCD critical point}
\label{sec:SDEc}

In this appendix, we explain that the appropriate equation to describe 
the time evolution of the soft mode near the critical point is 
the SDE~(\ref{eq:SDE}), on the basis of 
Refs.~\cite{Hohenberg:1977ym,Fujii:2004jt,Son:2004iv}.

At sufficiently long distance and time scales, the dynamics of a 
finite temperature system near equilibrium is described by hydrodynamic theory, 
which only contains the modes whose excitation energies vanish 
in the long wavelength limit -- ``hydrodynamic modes''.
Near the critical point, the hydrodynamic variables are given by 
the fluctuations of the order parameter and 
the densities of conserved charges~\cite{Fujii:2004jt,Son:2004iv}.
The QCD critical point shares the same dynamical universality class with
the model H in the classification of 
Ref.~\cite{Hohenberg:1977ym}, and the chiral order parameter field 
$\sigma=\langle\bar{q}q\rangle$ has nonzero couplings with 
baryon number density $n$, and energy and momentum densities.
In this appendix, we neglect the energy-momentum density 
for a simple illustration.

We start from the Ginzburg-Landau free energy functional
\begin{align}
\label{eq:GL}
&F[\sigma(\bm{x}),n(\bm{x})]
\nonumber \\
&=
\frac12 \int d{\bm x}\left[
A (\delta\sigma)^2 + 2B(\delta\sigma)(\delta n)
+ C (\delta n)^2 + \cdots
\right],
\end{align}
where the coefficients $A,\ B$, and $ C$ are functions of temperature $T$ 
and baryon chemical potential $\mu$. 
The neglected terms include higher order terms in $\delta\sigma$ and 
$\delta n$, and derivative terms, which are not important 
to describe slow modes.
Here, $B\neq0$ because the coupling between $\sigma$ and $n$ is 
allowed at the QCD critical point because of the finite quark 
masses and finite baryon density~\cite{Fujii:2004jt}.

Deviation of $\sigma$ and $n$ from the equilibrium values gives 
rise to relaxation of the system.
The evolutions of $\sigma$ and $n$ are given by the following 
stochastic hydrodynamic equations:
\begin{eqnarray}
\label{eq:TDGL}
\left(
\begin{array}{c}
\dot{\sigma} \\
\dot{n}
\end{array}
\right)
= -\left(
\begin{array}{cc}
\gamma_{\sigma\sigma} & \gamma_{\sigma n}  \\
\gamma_{n\sigma} & \gamma_{nn}
\end{array}
\right)
\left(
\begin{array}{c}
\frac{\delta F}{\delta \sigma} \\
\frac{\delta F}{\delta n}
\end{array}
\right) +\left(
\begin{array}{c}
\xi_{\sigma} \\
\xi_{n}
\end{array}
\right),
\end{eqnarray}
where the noise correlators are local,
\begin{eqnarray}
\langle \xi_i({\bm x}_1,t_1)\xi_j({\bm x}_2,t_2) \rangle_{\rm c}
\sim \delta({\bm x}_1-{\bm x}_2)\delta(t_1-t_2),\quad
\end{eqnarray}
with $i,j=\sigma,n$.
From Onsager's principle, we have $\gamma_{\sigma n}=\gamma_{n\sigma}$.
In the small momentum limit, 
$\gamma_{\sigma\sigma}$ is given by a nonzero constant,
while the coefficients for $\dot{n}$, 
$\gamma_{n\sigma}$ and $\gamma_{nn}$, are proportional to space-derivative
squared because of the conservation law, parity invariance, and
analyticity. We thus have in the Fourier space 
\begin{align}
\gamma_{\sigma n} = \gamma_{n\sigma} = \tilde\lambda q^2, \quad 
\gamma_{nn} = \lambda q^2,
\label{eq:gamma}
\end{align}
with $q^2$ being the momentum squared.
Inserting Eqs.~(\ref{eq:gamma}) and (\ref{eq:GL}) into Eq.~(\ref{eq:TDGL}), 
we obtain the hydrodynamic equation to leading order in $q^2$ as
\begin{align}
\label{eq:hydro}
\left(
\begin{array}{c}
\dot{\sigma} \\
\dot{n}
\end{array}
\right)
=& -\left(
\begin{array}{cc}
\gamma_{\sigma\sigma} A & \gamma_{\sigma\sigma} B  \\
(\tilde{\lambda} A+\lambda B) q^2 & 
(\tilde{\lambda} B+\lambda C) q^2
\end{array}
\right)
\left(
\begin{array}{c}
\sigma \\
n
\end{array}
\right)
\nonumber \\
&+\left(
\begin{array}{c}
\xi_{\sigma} \\
\xi_{n}
\end{array}
\right).
\end{align}

Solving Eq.~(\ref{eq:hydro}), we obtain two eigenfrequencies,
\begin{eqnarray}
\label{eq:1}
\omega_1 = -i\lambda\frac{\Delta}{A}q^2, 
\quad
\omega_2 = -i\gamma_{\sigma\sigma} A , 
\end{eqnarray}
with $\Delta=AC-B^2$ and 
the corresponding eigenmodes,
\begin{eqnarray}
v_1 = \delta n ,
\quad
v_2 = A \delta \sigma + B \delta n
,
\label{eq:2}
\end{eqnarray}
which decay with $|\omega_1|$ and $|\omega_2|$, respectively.
At the critical point, the energy functional develops a flat direction,
$\Delta=0$, and the susceptibilities of $\sigma$ and $n$ become divergent.
The diffusion coefficient
\begin{eqnarray}
D=\lambda \frac{\Delta}{A},
\end{eqnarray}
goes to zero at the critical point, which 
represents the critical slowing down.

This result tells us that 
a small fluctuation of $\sigma$ and $n$ in the system 
relaxes with two distinct time scales. 
The first slow mode $v_1$ is just the conserved charge $n$, 
whose time evolution is described by the SDE
\begin{eqnarray}
\label{eq:SDEc}
\dot{v}_1 = D \nabla^2 v_1 + \xi_n .
\end{eqnarray}
Note that $\xi_n$ should be proportional to space-derivative as 
in Eq.~(\ref{eq:SDE})
so that this equation is consistent with the continuity equation.
On the other hand, $v_2$ is a relaxation mode with nonvanishing 
relaxation time scale $(\gamma_{\sigma\sigma}A)^{-1}$.
This time scale is fast compared to that of $v_1=\delta n$.
In the faster time scale, $\sigma$ alone relaxes to the value 
$\delta \sigma=-(B/A)\delta n$, which 
minimize $F[\sigma,n]$ with $n$ being fixed to a given value.
After that, with the longer time scale $(Dq^2)^{-1}$ the mode $v_1$
relaxes to $\delta n=0$. In this stage, $\sigma$ simply traces the 
profile of $n$, and the time evolution of the slow mode is 
described by the SDE (\ref{eq:SDEc}).
The effect of the critical point is 
encoded in vanishing of $D$ in the SDE.

In the above discussion we neglected the energy-momentum density.
In model H, with which the QCD critical point shares the same dynamic universality, 
the hydrodynamic slow modes in the long wavelength limit are in fact 
given by the baryon number diffusion and the diffusion of two transverse momentum components~\cite{Son:2004iv} having a nonlinear coupling
via the nonzero Poisson bracket~\cite{Hohenberg:1977ym}.
The analysis incorporating the nonlinear coupling is left for future works.

\section{Conditions for the appearance of non-monotonicity}
\label{sec:discussion}

In Sec.~\ref{sec:analytic}, we showed that the non-monotonic behaviors of 
$K(\Delta y)$ and $C(\bar{y})$ in Eqs.~(\ref{eq:K}) and (\ref{eq:C}) 
serve as direct experimental evidence for the existence of 
a peak structure in $\chi(\tau)$.
In this appendix, we take a much closer look at the conditions for the 
appearance of the non-monotonic behaviors in these functions and discuss
which function is better in sustaining the non-monotonicity.

To simplify the problem, in this appendix we consider the functional
form of $\chi(\tau)$ which has only one maximum as a function of $\tau$.
Then, $\chi'(\tau)$ changes its sign only once from positive to negative.
In this case, $K(\Delta y)$ ($C(\bar{y})$) can have only one local 
maximum (minimum).
Thus, the necessary and sufficient conditions for the non-monotonic behaviors
of $K(\Delta y)$ and $C(\bar{y})$ are given by
\begin{align}
&\lim_{\Delta y\to0}
\frac{d K(\Delta y)}{d \Delta y}>0
 ~\mbox{ and }
\lim_{\Delta y\to\infty}
\frac{d K(\Delta y)}{d \Delta y}<0,
\label{eq:K><}
\\
&\lim_{\bar{y}\to0}
\frac{d C(\bar{y})}{d \bar{y}}<0 
~\mbox{ and }
\lim_{\bar{y}\to\infty}
\frac{d C(\bar{y})}{d \bar{y}}>0,
\label{eq:C><}
\end{align}
respectively.

From Eq.~(\ref{eq:cum}), the $\Delta y$ derivative of 
$K(\Delta y)$ is given by 
\begin{align}
\label{eq:K'}
\frac{d K(\Delta y)}{d \Delta y}
=- \int_{\tau_0}^{\tau_{\rm f}}d\tau' 
\frac{\chi^{\prime}(\tau')}{2\chi^{\rm H}d(\tau',\tau_{\rm f})}
F'\Big( \frac{\Delta y}{2d(\tau',\tau) } \Big),
\end{align}
where, $F'(X)=dF(X)/dX$. 
Using 
\begin{align}
\lim_{X\to0}F'(X)= \pi^{-1/2}, \quad 
\lim_{X\to\infty}F'(X)= \pi^{-1/2} X^{-2},
\label{eq:limF'}
\end{align}
one obtains
\begin{align}
\lim_{\Delta y\to0}
\frac{d K(\Delta y)}{d \Delta y}
=& 
-\frac1{2\sqrt{\pi}} \int_{\tau_0}^{\tau_{\rm f}}d\tau' 
\frac{\chi^{\prime}(\tau')}{\chi^{\rm H}d(\tau',\tau_{\rm f})},
\label{eq:K'->0}
\\
\lim_{\Delta y\to\infty}
\frac{d K(\Delta y)}{d \Delta y}
=& 
-\frac1{2\sqrt{\pi}(\Delta y)^2} \int_{\tau_0}^{\tau_{\rm f}}d\tau' 
\frac{\chi^{\prime}(\tau')d(\tau',\tau_{\rm f})}{\chi^{\rm H}}.
\label{eq:K'->inf}
\end{align}
Here, $d(\tau',\tau_{\rm f})$ is a monotonically decreasing function 
of $\tau'$ with $d(\tau_{\rm f},\tau_{\rm f})=0$.
The integrand in Eq.~(\ref{eq:K'->0}) is $\chi'(\tau')$ with 
a weight $1/d(\tau',\tau_{\rm f})$, which takes larger value 
for larger $\tau'$.
The sign of $\chi'(\tau')$ with later $\tau'$ is more strongly reflected
to the sign of Eq.~(\ref{eq:K'->0}).
On the other hand, in Eq.~(\ref{eq:K'->inf}) $\chi'(\tau')$ is integrated
with a weight $d(\tau',\tau_{\rm f})$ taking larger value for earlier $\tau'$.
The sign of $\chi'(\tau')$ with earlier $\tau'$ is more responsible 
for that of Eq.~(\ref{eq:K'->inf}).

Next, the $\bar{y}$ derivative of $C(\bar{y})$ 
is calculated to be
\begin{align}
\frac{d C(\bar{y})}{d \bar{y}}
= 
\frac{\bar{y}}{4\sqrt{\pi}} \int_{\tau_0}^{\tau_{\rm f}}d\tau' 
\frac{\chi'(\tau')}{\chi^{\rm H}}
\frac{e^{-\bar{y}^2/d(\tau',\tau_{\rm f})^2}}{d(\tau',\tau_{\rm f})^3}.
\label{eq:C'}
\end{align}
By taking the small $\bar{y}$ limit, we obtain
\begin{align}
\lim_{\bar{y}\to0}
\frac{d C(\bar{y})}{d \bar{y}}
=& 
\frac{\bar{y}}{4\sqrt{\pi}} \int_{\tau_0}^{\tau_{\rm f}} d\tau' 
\frac{\chi'(\tau')}{\chi^{\rm H} d(\tau',\tau_{\rm f})^3}.
\label{eq:C'->0}
\end{align}
Equation~(\ref{eq:C'->0}) shows that the sign of 
$d C(\bar{y})/d \bar{y}$ in the small $\bar{y}$ limit
is determined by the integral of $\chi'(\tau')$ with a 
weight $1/d(\tau',\tau_{\rm f})^3$.
In the large $\bar{y}$ limit, the weight is
given by $e^{-\bar{y}^2/d(\tau',\tau_{\rm f})^2}/d(\tau',\tau_{\rm f})^3$, 
which concentrates at the initial time $\tau'=\tau_0$ 
in the large $\bar{y}$ limit.
The sign in this limit thus is determined only by
$\chi'(\tau_0)$, which is positive in the present situation.

Now, let us compare the conditions Eqs.~(\ref{eq:K><}) and (\ref{eq:C><}).
As discussed above, the second condition in Eq.~(\ref{eq:C><}) is 
always satisfied, while that in Eq.~(\ref{eq:K><}) is not necessarily
true but dependent on the functional form of $\chi(\tau)$.
Next, the first conditions in Eqs.~(\ref{eq:K><}) and (\ref{eq:C><})
are not always satisfied, but the latter is more favored,
because the weight of the latter, $d(\tau',\tau_{\rm f})^{-3}$, is 
more concentrated at later $\tau'$ than the former,
$d(\tau',\tau_{\rm f})^{-1}$.
From these observations, one concludes that the appearance of the 
non-monotonicity is more likely to appear in $C(\bar{y})$ 
than in $K(\Delta y)$.

Although the manifestation of non-monotonicity is more robust 
in $C(\bar{y})$ than $K(\Delta y)$, in experimental analyses
it is meaningful to analyze both of these functions.
In the above argument, the position of the local extremum 
is not determined.
The ranges of $\Delta y$ and $\bar{y}$ which can be measured
in experiments are limited owing to the coverage of the detector,
and the manifestation of the extremum in this range depends
on the functional form.
Therefore, by analyzing both $K(\Delta y)$ and $C(\bar{y})$,
the chance to find the non-monotonic behaviors is enhanced.

\end{document}